\documentclass[a4paper,11pt]{article}
\usepackage{jcappub} 

\title{\boldmath ANNZ+: an enhanced photometric redshift estimation algorithm with applications on the {PAU} {Survey}}

\author[a]{Imdad Mahmud~Pathi,}
\author[a]{John~Y.~H. Soo,}
\author[a]{Mao~Jie Wee,}
\author[a]{Sazatul~Nadhilah Zakaria,}
\author[a]{Nur~Azwin Ismail,}
\author[b]{Carlton~M. Baugh,}
\author[c]{Giorgio Manzoni,}
\author[d,e,f]{Enrique Gaztanaga,}
\author[e,f]{Francisco~J. Castander,}
\author[g,h]{Martin Eriksen,}
\author[h,i]{Jorge Carretero,}
\author[g]{Enrique Fernandez,}
\author[j]{Juan Garcia-Bellido,}
\author[g,k]{Ramon Miquel,}
\author[g]{Cristobal Padilla,}
\author[l]{Pablo Renard,}
\author[i]{Eusebio Sanchez,}
\author[i]{Ignacio Sevilla-Noarbe,} 
\author[h,i]{and Pau Tallada-Cresp\'{i}}

\affiliation[a]{School of Physics, Universiti Sains Malaysia, 11800 USM, Pulau Pinang, Malaysia}
\affiliation[b]{Institute for Computational Cosmology, Department of Physics, Science Laboratories, \\ Durham University, South Road, Durham, DH1 3LE, UK}
\affiliation[c]{Jockey Club Institute for Advanced Study, The Hong Kong University of Science and Technology, Hong Kong S.A.R., China}
\affiliation[d]{Institute of Cosmology and Gravitation (ICG), University of Portsmouth, PO1 3FX \\ Portsmouth, UK}
\affiliation[e]{Institute of Space Sciences (ICE-CSIC), Universitat Aut\`{o}noma de Barcelona, Carrer  de Can Magrans S/N, E-08193 Cerdanyola del Vall\'{e}s (Barcelona), Spain}
\affiliation[f]{Institut d'Estudis Espacials de Catalunya (IEEC), Edifici RDIT, Universitat Polit\'{e}cnica de Catalunya, E-08860 Castelldefels (Barcelona), Spain}
\affiliation[g]{Institut de F'{\i}sica d'Altes Energies (IFAE), Universitat Aut\`{o}noma de Barcelona, E-08193 Bellaterra (Barcelona), Spain}
\affiliation[h]{Port d'Informaci\'{o} Cient\'{i}fica (PIC), Universitat Aut\`{o}noma de Barcelona, Carrer Albareda S/N, E-08193 \\ Bellaterra (Barcelona), Spain}
\affiliation[i]{Centro de Investigaciones Energ\'eticas, Medioambientales y Tecnol\'ogicas (CIEMAT), Avenida Complutense 40, E-28040 Madrid, Spain}
\affiliation[j]{Instituto de F\'{i}sica Te\'{o}rica UAM-CSIC, Universidad Auton\'{o}ma de Madrid, E-28049 Cantoblanco (Madrid), Spain}
\affiliation[k]{Instituci\`{o} Catalana de Rercerca i Estudis Avan\c{c}ats (ICREA), Pg.~de Llu'{\i}s Companys 23, E-08010 Barcelona, Spain}
\affiliation[l]{Department of Astronomy, Tsinghua University, Beijing 100084, China}

\emailAdd{johnsooyh@usm.my}

\abstract{\textsc{annz} is a fast and simple algorithm which utilises artificial neural networks (ANNs), it was known as one of the pioneers of machine learning approaches to photometric redshift estimation decades ago. We enhanced the algorithm by introducing new activation functions like tanh, softplus, SiLU, Mish and ReLU variants; its new performance is then vigorously tested on legacy samples like the Luminous Red Galaxy (LRG) and Stripe-82 samples from SDSS, as well as modern galaxy samples like the Physics of the Accelerating Universe Survey (PAUS). This work focuses on testing the robustness of activation functions with respect to the choice of ANN architectures, particularly on its depth and width, in the context of galaxy photometric redshift estimation. Our upgraded algorithm, which we named \textsc{annz+}, shows that the tanh and Leaky ReLU activation functions provide more consistent and stable results across deeper and wider architectures with $>1$ per cent improvement in root-mean-square error ($\sigma_{\textrm{RMS}}$) and 68th percentile error ($\sigma_{68}$) when tested on SDSS data sets. While assessing its capabilities in handling high dimensional inputs, we achieved an improvement of 11 per cent in $\sigma_{\textrm{RMS}}$ and 6 per cent in $\sigma_{68}$ with the tanh activation function when tested on the $40$-narrowband PAUS dataset; it even outperformed \textsc{annz2}, its supposed successor, by 44 per cent in $\sigma_{\textrm{RMS}}$. This justifies the effort to upgrade the 20-year-old \textsc{annz}, allowing it to remain viable and competitive within the photo-$z$ community today. The updated algorithm \textsc{annz+} is publicly available at \url{https://github.com/imdadmpt/ANNzPlus}.}

\begin{document}
\maketitle
\flushbottom
\section{Introduction}\label{sec:intro}
In recent years, the astronomical community has faced a data deluge reaching exabytes due to the enormous efforts to comprehend the cosmos \citep{rodriguez_application_2022}. Numerous planned and ongoing surveys such as the Sloan Digital Sky Survey (hereafter, SDSS) \cite{york_sloan_2000}, Dark Energy Survey (DES) \cite{collaboration_dark_2005}, \textit{Euclid} \cite{laureijs_euclid_2011}, Kilo-Degree Survey (KiDS) \cite{de_jong_kilo-degree_2013}, Nancy Grace Roman Space Telescope (formerly Wide-Field Infrared Survey Telescope or WFIRST) \cite{spergel_wide-field_2013}, Dark Energy Spectroscopic Instrument (hereafter, DESI) \cite{levi_DESI_2013}, Subaru Prime Focus Spectrograph (PFS) \cite{tanaka_PFS_2014}, Hyper Suprime-Cam (HSC) \cite{aihara_hyper_2017}, Legacy Survey of Space and Time (LSST) \cite{ivezic_lsst_2019}, and the 4-meter Multi-Object Spectroscopic Telescope (4MOST) \cite{dejong_4most_2019} are expected to observe a significant number of galaxies in both wide and deep areas. These surveys will be able to provide highly accurate measurements of the large-scale structure of the Universe, a better understanding of the formation and evolution of the galaxies changed over time and the nature of dark energy and dark matter. Additionally, brand-new resources like the \textit{James Webb Space Telescope} (hereafter, \textit{JWST}) \cite{yan_first_2023} allowed the first batch of galaxies with $11 < z < 20$ to be observed. The latest discovery of early galaxies, using \textit{JWST} medium band filters \cite{withers_emmisionline_2023}, highlights the crucial role of photometric redshifts in the study of the Universe's structure, and in providing candidates for follow-up spectroscopy. These surveys require good quality photometric redshifts (photo-$z$s) to attain their research targets. 

Spectroscopic redshifts, or spec-$z$s refer to the displacement of spectral lines measured by a spectrograph, primarily due to the expansion of the Universe. These redshifts are generally determined by analysing the galaxy's spectrum, but the process requires a lot of telescope time and uses expensive facilities. Even with the advent of spectrographs with a huge multiplex factor, like DESI which can target 5,000 objects in one pointing, deeper observations can reach much higher surface densities of galaxies than this, requiring many passes of the same area to achieve high spectroscopic completeness.  
This is where photo-$z$s come in: photo-$z$ is a technique that allows us to estimate the distance to a galaxy based on its observed photometry. It is called "photometric" because it involves measuring the amount of light emitted by a galaxy at different wavelengths. After all, the observed wavelengths of light are typically redshifted, or shifted to longer wavelengths, as the light travels through expanding space \citep{koo_optical_1985,loh_photometric_1986}. Photo-$z$s are usually estimated using broad-band magnitudes and there are two main methods: spectral template fitting and empirical/machine learning techniques. The first approach commonly employs a likelihood utilising a $\chi^2$ fit including codes like \textsc{delight} \citep{leistedt_hierarchical_2022}, \textsc{gazelle} \citep{kotulla_impact_2009}, \textsc{eazy} \citep{brammer_eazy_2008}, \textsc{zebra} \citep{feldmann_zurich_2006}, \textsc{hyperz} \citep{bolzonella_photometric_2000}, \textsc{bpz} \citep{benitez_bayesian_2000} and \textsc{le phare} \citep{arnouts_measuring_1999}. The second approach employs machine learning methodologies, like self-organizing maps \citep{bunchs_self_organizing_2019}, neural networks \citep{lee_estimation_2021}, deep convolutional neural networks \citep{pasquet_photometric_2019}, galaxy morphoto-$z$ with neural networks (\textsc{gaznets}) \cite{li_galaxy_2022}, multi-task learning networks \citep{cabayol_multitask_2023}, interpretable deep capsule networks \citep{dey_photometric_2022} and bayesian neural networks \citep{jones_photometric_2023}. 

Artificial Neural Network Redshift (\textsc{annz}\footnote{\url{http://www.homepages.ucl.ac.uk/~ucapola/annz.html}}) introduced by \cite{collister_annz_2004} was among the first machine learning photometric redshift algorithms produced and widely used; it was used for photo-$z$ estimation \citep{collister_megaz-lrg_2007,bonfield_photometric_2010} concurrently with other photo-$z$ codes \citep{hildebrandt_phat_2010,abdalla_comparison_2011,reis_sloan_2012,sanchez_photometric_2014,bundy_galaxy_2015,soo_morpho-z_2018}. It has also been employed for morphological classification of galaxies \citep{lahav_neural_1996,ball_galaxy_2004,banerji_photometric_2008} and star-galaxy separation \citep{bertin_sextractor_1996,soumagnac_galaxy_2015,kim_classification_2016}. \textsc{annz} has been replaced by its successor \textsc{annz2}\footnote{\url{https://github.com/IftachSadeh/ANNZ}} \citep{sadeh_annz2_2016} which includes some quantification of epistemic uncertainties through an ensemble of randomised estimator techniques reminiscent of modern deep ensembles. Despite the latest photo-$z$ codes, \textsc{annz} remains fast and simple compared to many other algorithms.

In recent work, \textsc{annz} was used in ref. \cite{collaboration_euclid_2020} as one of the 13 photometric redshift codes to assess the strengths and weaknesses of current photo-$z$s methods for \textit{Euclid} mission focusing particularly on the $0.2 - 2.6$ redshift range that the \textit{Euclid} mission will probe in order to provide precise and accurate photo-$z$ measurements. The results showed \textsc{annz} was able to provide reliable single value estimates for redshifts but they faced challenges in producing useful PDF for the validation sample sources. \textsc{annz} also experienced sharp drops in performance metrics above redshift $z \approx 1.3$ which it struggled with maintaining performance at higher redshifts but \textsc{annz} showed good results and performed generally better than the template-fitting ones in the lower redshift. Despite various machine learning algorithms, after decades, \textsc{annz} is still in use today.

However, ref. \cite{soo_pau_2021} demonstrated that artificial neural networks can yield poor results when used with high-dimensional narrowband inputs without comprehensive optimisation. Ref. \cite{collaboration_euclid_2022} mentioned they have not fully explored all the possible combinations of the number of neurons and hidden layers using DLNN and CNN. This is the opportunity of this work to provide insight since \textsc{annz} has the ability to do it. 
The significant recent advances in the technology and theory of artificial neural networks raise the question: is \textsc{annz} still competitive 20 years after its debut?

One of the most critical hyperparameters affecting the performance of an artificial neural network (hereafter, ANN) is the choice of activation function. Activation functions are mathematical functions that determine the output layer of a neuron in an ANN and transform the weighted input of a neuron into more linearly separable abstract features using multiple layers, enhancing the effectiveness of the ANN by creating non-linear relationships between variables. The most popular activation functions are the sigmoid \cite{Grossberg_brain_1982}, hyperbolic tangent (tanh) \cite{lecun_backpropagation_1989}, softplus \cite{dugas_incorporating_nodate_2000}, rectified linear unit (ReLU) \cite{Nair_relu_2010}, leaky rectified linear unit (Leaky ReLU) \cite{maas_leakyrelu_2013}, exponential linear unit (ELU) \cite{clevert_fast_2015}, parametric rectified linear unit (PReLU) \cite{he_prelu_2015}, S-shaped rectified linear unit (SReLU) \cite{jin_deep_2016}, sigmoid linear unit (SiLU) \cite{hendrycks_gaussian_2016}, and mish \cite{misra_mish_2019}. Activation functions can be categorised into three groups, which are fixed, adaptive and nonstandard activation functions \citep{dubey_surveyAF_2022,jagtap_activation_2023}. Despite using common trial and error methods to optimise ANN's hyperparameters and test various activation functions, ref. \cite{santos_gainn_2024} did not draw any conclusions on which activation functions should be used.

Since \textsc{annz} debuted way before many of these activation functions were introduced, it only utilised one single activation function in its code: the logistic sigmoid. We were interested in testing the viability and competitiveness of \textsc{annz} by "updating" it with the choice of these more up-to-date activation functions while analysing their performance as well. \textsc{annz} is a machine learning algorithm coded from scratch using C++, therefore it gives us the flexibility to insert and test any known activation function in the literature, and is not limited to those made available in public machine learning packages. We believe that the study of activation functions in the context of regression can significantly impact the accuracy and precision of galaxy photo-$z$s of modern surveys, which currently still relies heavily on the synergy of results between many different algorithms and methodologies. 

Therefore, we will address the following objectives in this work:

\begin{enumerate}
    \item To improve the accuracy of photometric redshift estimation by identifying stable and consistent activation functions across varying architectures of artificial neural networks, and therefore assess the relevancy and competitiveness of our updated algorithm \textsc{annz+};
    \item To gain valuable insights into how different activation functions respond to the complexity of high dimensional input, in the context of photometric redshift estimation.
\end{enumerate}

While many may deem the change in activation function as elementary or merely one of the many ways to optimise ANNs, we emphasise that this work utilises this simple methodology to kill two birds in one stone: to improve the original \textsc{annz}, and to break the myth that 'ReLU is the best', which is too common a suggestion found on many online resources in regards to deep learning. We believe that the context is very important: in this case, which is of regression and photo-$z$, we specifically chose to study the impact of activation function in regards to this specific context while keeping all other hyperparameters constant. It is hoped that this work would act as a detailed report with quantitative evidence on ANN optimisation: a resource which is relevant and significant within the current photo-$z$ community.

This paper is structured as follows. In section 2, we first introduce the algorithm, activation functions and metrics used. Section 3 provides details on the data sets and training samples used. We present answers and discussions for objectives (1) and (2) in section 4, followed by a continued discussion of objective (1) in section 5. Our work concludes in section 6 with suggestions for future work. This work is a continuation and completion of the preliminary work presented in a proceeding by \cite{soo_ml_2023}.

\section{Methodology}\label{sec:method}
This study aims to enhance existing machine learning techniques for photo-$z$ estimation by choosing the suitable activation function. The machine learning algorithm discovers a deterministic relationship between the input variables (e.g. the $ugriz$ broadband magnitudes) and the spectroscopic redshifts in a training set (under the assumption that the spectroscopic redshift is the true redshift) and then uses this information to produce photo-$z$s for a target sample without any available spectroscopic redshift information. We will briefly review \textsc{annz} in the subsection below. Then, we will list the activation functions explored in section~\ref{sec:activation}, followed by a description of the performance metrics used in this study in section~\ref{sec:metrics}.

\subsection{ANNZ}\label{sec:annz}

\begin{figure}
\includegraphics[width=\linewidth]{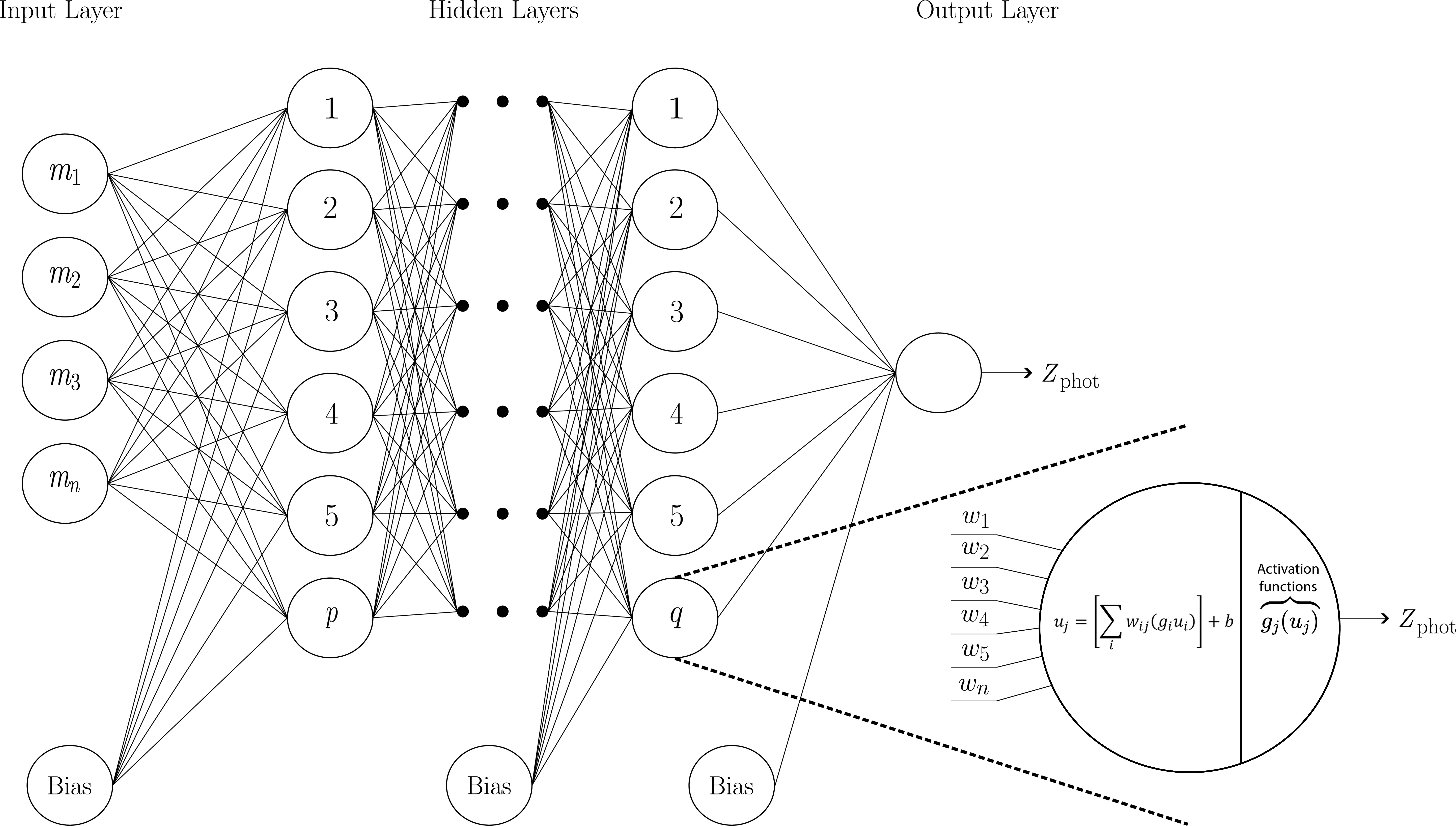}
\caption{A schematic diagram of the complex inner workings of deep artificial neural networks (ANNs) with the activation function $g_{j} (u_{j})$ embedded in each node, highlighting their crucial role in shaping the network's behaviour and facilitating the production of efficient photo-$z$ information. The variables $m_1,...,m_n$ represent the inputs, $w_{ij}$ denote the weights, the circles symbolise the neurons, and every vertical set of circles is a layer. The number of nodes in a layer is labelled as $p$ and $q$ while the output is denoted as $z_{\text{phot}}$. The bias node $b$ allows the function to be shifted up or down.} 
\label{fig:ANN}
\end{figure}

Artificial Neural Network redshift estimation algorithm, known as \textsc{annz} \cite{collister_annz_2004} has been used to produce many photo-$z$ catalogues \citep{abdalla_predicting_2008,banerji_photometric_2008}. \textsc{annz} is a multilayer perceptron (MLP) comprised of multiple layers of interconnected nodes. The input layer of nodes is the galaxy magnitudes in the available filters $m = (m_{1}, m_{2},..., m_{n})$. The final layer is the single output node, the photometric redshift $z_{\textrm{phot}}$. The intermediate layers are called “hidden layers”, the size and number of which are entirely up to the user. The network architecture is represented as $N_{\textrm{in}}: N_{1}: N_{2}:...: N_{\textrm{out}}$, where $N_{\textrm{in}}$ is the number of input nodes, followed by $N_{1}$ in the first hidden layer, and subsequently, the number of nodes in the $i{\text{th}}$ intermediate hidden layers are denoted by $N_{i}$. The neural network architecture 5:10:10:1 is an example of 5 input nodes, 10 nodes in each of the two hidden layers and one output node. Each interconnecting link between the nodes is associated with a weight $w_{ij}$ to be optimised, where $i$ and $j$ describe the two nodes. An activation function, $g_{j} (u_{j})$, which introduce non-linearity to the model is applied at each node $j$ and is defined as:
\begin{equation}
\label{eqn:acivation} {u_{j}} = \left[\sum_{i}w_{ij}g_{i}(u_{i})\right] + b \ ,  
\end{equation}
where the node $u_{j}$ is the value at node $j$ that carries the weighted sum over all nodes $u_{i}$ to nodes $j$ and $g_{i}(u_{i})$ is the activation function applied to the value $u_{i}$ of node $i$. The bias node, $b$, allows the function to be shifted up or down, which can be used to adjust the accuracy of the function. Figure~\ref{fig:ANN} represents the building blocks of a deep ANN, which emphasises neurons producing an output by combining the linear combination of inputs and then applying a non-linear transformation known as the activation function. The original activation function used in \textsc{annz} is the sigmoid function.

The galaxy training set needs to have photometry, $m$, and spectroscopic redshift, $z_{\textrm{spec}}$, to train the ANN by adjusting the weights to minimise the cost function. The square in the cost function gives more weight to the outliers rather than taking the absolute value.
\begin{equation}
\label{eqn:cost} E = \sum_{k}[z_{\textrm{phot}}(w,m_{k}) - z_{\textrm{spec}, \textit{k}}]^2.   
\end{equation}
To make sure the weights are \textit{regularised},  an extra quadratic cost term 
\begin{equation}
\label{eqn:extra} E = \beta\sum_{i,j} w^2_{ij}, 
\end{equation}
is added to eq.~(\ref{eqn:cost}): this term keeps the weight values small to avoid overfitting. 

An iterative quasi-Newtonian method is used in \textsc{annz} to execute the minimisation process (refer to \cite{bishop_neural_1995,lahav_neural_1996} for more details). The cost function is evaluated on a validation set after each iteration. Training stops when the cost function is minimised on the validation set to prevent overfitting. The final weights from this iteration are chosen for the ANN, which can then be used to process new input vectors and compute the outputs.

For photo-$z$, the network architecture of 5:10:10:1 was found empirically to be optimal \citep{firth_estimating_2003,collister_megaz-lrg_2007,abdalla_comparison_2011} for MegaZ-LRG. We have set this architecture as the default setting proposed by \cite{collister_annz_2004}. In this study, we compare the performance of this default setting with two additional types of networks:

\begin{enumerate}
    \item a deep network, denoted as $N_{\textrm{in}}:10_{d}:1$, where $d$ is the number of hidden layers, e.g. for $d = 2$, we have $N_{\textrm{in}}:10:10:1$, $d = 3$, we have $N_{\textrm{in}}:10:10:10:1$ and etc. We tested using up to ten hidden layers, with each layer consisting of ten nodes.
    
    \item a wide network, with a hierarchical structure denoted as $N_{\textrm{in}}:10N:10N:1$, where $N$ represents an increasing number of nodes ranging from one to ten. For instance, if $N = 5$, the network would have 50 nodes in the first and second hidden layers (i.e., $10 \times 5$), giving an architecture configuration of 5:50:50:1.
\end{enumerate}

The method for choosing the random seed is typically via trial and error. Each committee's (or network's) training with different random seeds enhances the diversity of the learning process, thereby reducing the chances of overfitting. This approach ensures our model can generalise effectively when faced with new and unseen data. The different initial random seeds may lead to similar RMS accuracy, but slight differences of around $\lesssim 10$ per cent can still exist \cite{firth_estimating_2003}. However, here, some activation functions may influence the consistency of the RMS accuracy (see section 4.2 for a further discussion on this). 

We have chosen the default setting of the ANN to run up to a maximum of 9999 iterations; four committees were trained for each network architecture, and each was initialised with four fixed random seeds with values of [5, 50, 500, 5000]. Unfortunately, due to the randomised initialisation of the weights, two scenarios occasionally occurred: (1) \textsc{annz} encountered operational limitations 'NaN' and (2) \textsc{annz} stopped early during the training, therefore in these situations, different random seeds will be utilised. 

\subsection{Activation functions}\label{sec:activation}

Activation functions play an important role in determining the transformation of inputs and outputs within each layer of a neural network. However, under certain conditions, some activation functions may cause problems that significantly affect the network's performance and the stability of its training process. The most common problems include:

\begin{enumerate}
    \item The \textit{vanishing gradient} problem, which occurs when the gradients of the loss function (which are used to update the weights of the neural network during backpropagation) become very small (close to zero) as they are propagated backwards through the layers and diminish. As a result, the network has almost no gradient to update the weights and enters into a saturated region, hindering effective learning \cite{Hochreiter_vanishing_1998}. The sigmoid and tanh functions are commonly associated with this problem. 

    \item The \textit{exploding gradient} problem, which arises due to poor weight initialisation or when the gradients of the loss function become extremely large as it propagates backwards through layers, causing the weight to be updated disproportionately, leading the network to diverge or produce 'NaN' values and fail to learn anything useful \cite{Pascanu_difficulty_2013}. Our results showed that the Softplus and Mish activation functions experience these issues when applied within complex network architectures. 
    
    \item The \textit{dying gradient} problem, which refers to a situation where certain neurons stop learning altogether, resulting in “dead" neurons that contribute nothing to the model’s output. This typically occurs with activation functions like ReLU, which produces zero output for all negative inputs. Consequently, the gradients of the loss function become zero during backpropagation, causing the weights to remain unchanged and preventing any further learning \cite{glorot_sparse_2011}.
\end{enumerate}

\begin{figure}
    \includegraphics[width=\linewidth]{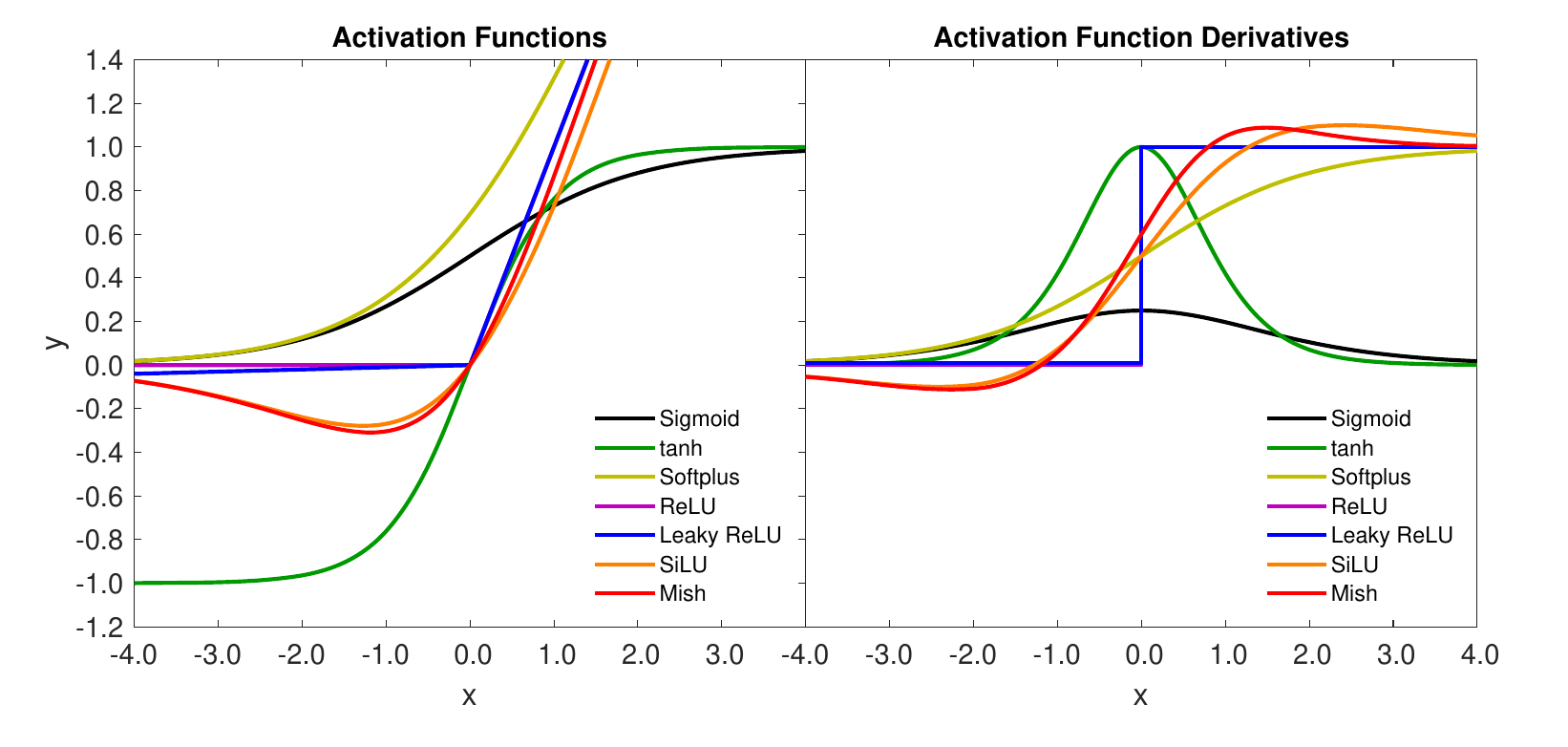}
    \caption{The activation functions and their derivatives: sigmoid, tanh, softplus, ReLU, Leaky ReLU, SiLU and Mish. Each function is represented by a different colour: black, green, yellow, magenta, blue, orange and red, as indicated by the key.}
    \label{fig: activation functions}
\end{figure}

By understanding these issues upfront, we can analyse the impact of the activation functions on the performance of \textsc{annz}. Here, we summarise and review several activation functions as potential replacements for the sigmoid function originally used in \textsc{annz}: tanh, SiLU, softplus, ReLU, Mish, and Leaky ReLU for the benefit of photometric redshift estimation. The shapes and ranges of the activation functions are depicted in figure~\ref{fig: activation functions} and to encapsulate the essence of activation functions, a summary can be found in table~\ref{tab:active summary}. 

\begin{table}
        \begin{tabular}{p{12mm} p{44mm} p{66mm} p{15mm}} 
        \hline
        \textbf{Name} & \textbf{Equation} & \textbf{Derivative} & \textbf{Range}\\ 
        \hline \\
        Sigmoid & $g(u) = \sigma(u) = \dfrac{1}{1+e^{-u}}$ & $g'(u) = g(u)[1 - g(u)]$ & (0,1)\\
        \\
        tanh & $g(u) = {\textrm{tanh}}(u)$ & $g'(u) = 1 - g(u)^2$ & (-1,1)\\
        \\
        Softplus & $g(u) = {\textrm{ln}}(1+e^{u})$ & $g'(u) = \sigma(u)$ & $(0,\infty)$\\
        \\
        ReLU & $g(u) = {\textrm{max}}(0,u)$ & 
        $g'(u) =   
        \begin{cases} 
            0, & u < 0 \\
            1, & u \geq 0 
        \end{cases}$ & $[0,\infty)$ \\
        \\ Leaky ReLU & $g(u) = {\textrm{max}}(0.01u,u)$ &
        $g'(u) =   
         \begin{cases} 
            0.01, & u < 0 \\
             1, & u \geq 0 
         \end{cases}$ & $(\infty,-\infty)$ \\
         \\
         SiLU & $g(u) = u \cdot \sigma(u)$ & $g'(u) = \sigma(u)[1+u[1-\sigma(u)]]$ & $[-0.28,\infty)$ \\
         \\
         Mish & $g(u) = u \cdot {\textrm{tanh}}[{\textrm{softplus}}(u)]$ & $g'(u) = {\textrm{sech}}^2[\textrm{softplus}(u)] \cdot \textrm{silu} (u) + \dfrac{g(u)}{u} $ & $[-0.31,\infty)$ \\ \\
        \hline
        \end{tabular}
        \centering
	\caption{Summary of the activation functions, showing their functional forms, derivatives and range.}  
	\label{tab:active summary}
        \centering
\end{table} 

\subsubsection{Logistic Sigmoid}\label{sec:sigmoid}

The logistic sigmoid activation function or also known as the logistic function \cite{Russell_AI_1995}
(hereafter, it will be referred to as sigmoid, which is more conventional) was used in ANN by Grossberg in 1982 \cite{Grossberg_brain_1982} and was the most popular activation function in the early 1990s \cite{Hecht_theory_1992}. It is defined as 
\begin{equation}
\label{eqn:sigmoid} g_{j}(u_{j}) = \frac{1}{1+e^{-u_{j}}} \ ,
\end{equation}
with derivative
\begin{equation}
\label{eqn:d_sigmoid} g_{j}'(u_{j}) = g_{j}(u_{j})[1 - g_{j}(u_{j})] \ .   
\end{equation}
The function takes any real number as input and produces an output that is also a real number. As shown in figure~\ref{fig: activation functions}, the function is differentiable and bounded to a range between 0 and 1 except for the endpoints themselves. The derivative of the sigmoid function is always positive, which makes it useful for backpropagation learning in neural networks \cite{Han_sigmoid_1995}. 

As mentioned in \ref{sec:annz}, \textsc{annz} used the sigmoid function as its default setting; however, this activation function faces significant problems during backpropagation. The vanishing gradient problem occurs when the outputs of the sigmoid function are very close to 1 or 0, causing the function to saturate when the inputs are large, either positively or negatively. Furthermore, the derivative of the sigmoid function is capped at a maximum of 0.25 (when the input is 0) and decreases rapidly as the input moves away from 0. 

Additionally, the output range of the sigmoid function (0, 1) is not centred at 0, which can lead to poor convergence \cite{Cui_NIR_2018}.  Moreover, using exponential functions in the sigmoid can increase computational demands, which might slow down the overall performance of the network. While the sigmoid activation function is useful for certain tasks like classification \cite{Hinton_speechrecognition_2012}, its drawbacks of vanishing and exploding gradient issues become significant in regression tasks (as evidenced by our findings) along with the limited output range and computational overhead, making it less desirable for more complex neural network architectures and tasks. It is also less effective when used in deeper networks \cite{Nair_relu_2010}. 

\subsubsection{Tanh}\label{sec:tanh}

The hyperbolic tangent (tanh) activation function \cite{lecun_backpropagation_1989} defined as
\begin{equation}
\label{eqn:tanh} g_{j}(u_{j}) = \frac{e^{u_{j}} - e^{-u_{j}}}{e^{u_{j}} + e^{-u_{j}}} = 2 \text{sigmoid}(2{u_{j}})-1 \ , 
\end{equation}
with derivative 
\begin{equation}
\label{eqn:d_tanh} g_{j}'(u_{j}) = 1 - g_{j}(u_{j})^2 \ , 
\end{equation}
which is the preferred choice over the sigmoid even though they share similarities to mapping input values to an S-shaped curve, as depicted in figure~\ref{fig: activation functions}. Besides, the tanh function has a wider range (-1, 1) than the sigmoid and it is centred at 0. Hence, this choice resolved one of the problems encountered with the sigmoid function. Due to this, the tanh function was frequently used in regression problems. However, the tanh function still faces saturated regions, and it remains exponential, which can introduce computational complexities.

\subsubsection{Softplus}\label{sec:softplus}

The softplus activation function \cite{dugas_incorporating_nodate_2000} serves as the primitive form of the sigmoid function and is mathematically defined as 
\begin{equation}
\label{eqn:softplus} g_{j}(u_{j}) = {\textrm{ln}}(1+e^{u_{j}}) \ ,   
\end{equation}
with derivative
\begin{equation}
\label{eqn:d_softplus} g_{j}'(u_{j}) = \frac{1}{1+e^{-u_{j}}} \ . 
\end{equation}
This is a convex function and has a non-saturating nature within the positive domain \cite{dugas_incorporating_nodate_2000}. The graph of the softplus function closely resembles ReLU (see next subsection), as depicted in figure~\ref{fig: activation functions}. Unlike the ReLU function, the softplus function exhibits slight reservations for values below 0, which can reduce the likelihood of neuronal degradation. However, it involves more computational complexity compared to the ReLU function \cite{wang_activation_2020}. An advantage of the softplus function lies in its smooth derivative, which proves beneficial for the backpropagation process in neural networks. Remarkably, the derivative of the softplus function is the sigmoid function. Notably, even with the ReLU function's zero-based threshold, neural networks trained using it tend to converge to local minima of equal or superior quality \citep{glorot_sparse_2011}.

\subsubsection{Rectified Linear Unit (ReLU)}\label{sec:relu}
ReLU \cite{glorot_sparse_2011} is a simple function,
\begin{equation}
\label{eqn:relu} 
g_{j}(u_{j}) = 
    \begin{cases} 
    0, & u_{j} < 0 \\
    u_{j}, & u_{j} \geq 0 \ ,
    \end{cases}
\end{equation}
with  derivative
\begin{equation}
g_{j}'(u_{j}) =   
    \begin{cases} 
    0, & u_{j} < 0 \\
    1, & u_{j} \geq 0 \ .
    \end{cases}
\end{equation}

It is a widely used activation function for many types of neural networks. ReLU substitutes negative values with 0 while leaving positive values unchanged. It converges faster than those using sigmoid or tanh activation functions because ReLU is non-saturating. It does not become saturated at high input values like sigmoid and tanh, thereby effectively alleviating the vanishing gradient problem during backpropagation \citep{Nair_relu_2010,glorot_sparse_2011}. This allows ReLU to be able to learn six times faster in large neural networks with large datasets than saturating non-linear activation functions \citep{Krizhevsky_NIPS_2012,Dahl_relu_2013,Montufar_linear_2014}.

However, ReLU is not centred at 0 and does not handle negative inputs well. This can cause problems during training. Leaky ReLU (section \ref{sec:Leaky ReLU}) is an extension of ReLU that addresses these issues.

\subsubsection{Leaky ReLU}\label{sec:Leaky ReLU}

Leaky ReLU \cite{maas_leakyrelu_2013} is an improved version of ReLU, specifically designed to address the dying ReLU problem (neuron weights not changing) and has all the advantages of ReLU. Leaky ReLU is defined as  
\begin{equation}
\label{eqn:leakyrelu} 
g_{j}(u_{j}) = 
    \begin{cases} 
    \alpha{u_{j}}, & u_{j} < 0 \\
    u_{j}, & u_{j} \geq 0 \ ,
    \end{cases}
\end{equation}
with derivative
\begin{equation}
\label{eqn:d_leakyrelu}
g_{j}'(u_{j}) =   
    \begin{cases} 
    \alpha, & u_{j} < 0 \\
    1, & u_{j} \geq 0 \ .
    \end{cases}
\end{equation}
The unique feature of Leaky ReLU is the addition of the parameter $\alpha$, which represents the small slope in the $u_{j} < 0$ region known as the leakage in the function. However, with this addition, the parameter $\alpha$ requires a value to be specified. In the majority of instances in literature, $\alpha$ = 0.01 is used \cite{jagtap_activation_2023}.

\subsubsection{Sigmoid Linear Unit (SiLU)}

SiLU was introduced by Hendrycks and Gimpel in 2016 \cite{hendrycks_gaussian_2016}. The function of SiLU is
\begin{equation}
    g_{j}(u_{j}) = u_{j}\cdot \Phi (u_{j}) = u_{j} \cdot \frac{1}{2} \left[1+ {\textrm{erf}} \left(\frac{u_{j}-\mu}{\sigma \sqrt{2}}\right) \right] \ ,
\end{equation}
where $\Phi(u_{j})$ represents the cumulative distribution function of a Gaussian distribution and erf stands for the error function and parameters $\mu = 0$ and $\sigma = 1$. SiLU can also be written as\\
\begin{equation}
    g_{j}(u_{j}) = u_{j} \cdot \sigma(u_{j}) = \frac{u_{j}}{1+ e^{-u_{j}}} \ ,
\end{equation}
and its derivative is,
\begin{equation}
    g_{j}'(u_{j}) =\frac{1+e^{-u_{j}} + u_{j}e^{-u_{j}}}{\left(1+e^{-u_{j}}\right)^2} \ ,
\end{equation}
where the sigmoid function $\sigma(u_{j}) = 1/\left(1 + e^{-u_{j}}\right)$ is used as an approximation to the cumulative distribution function of a Gaussian distribution.

Figure~\ref{fig: activation functions} shows SiLU are unbounded above and bounded below and distinguished by its smoothness and non-monotonicity. In fact, its non-monotonicity sets it apart from most typical activation functions as it has a global minimum value of $g_{j}(u_{j})\approx -0.28$ for $u_{j}\approx-1.28$. The global minimum, where the derivative is zero serves as an implicit regularisation that inhibits the learning of weights of large magnitudes \cite{elfwing_silu_2018}. Furthermore, being unbounded above aids in avoiding the saturation gradient problem and its smoothness also helps in the optimisation and generalisation of machine learning models \cite{hendrycks_gaussian_2016}. All these properties make SiLU perform much better than other activation functions in reinforcement learning. SiLU is also known as Swish or Sigmoid Shrinkage \citep{ramachandran_searching_2017,atto_smooth_2008}.

\subsubsection{Mish}\label{sec:mish}

Mish is derived from the author's name \cite{misra_mish_2019} and is defined as 

\begin{equation}
\label{eqn:mish} 
g_{j}(u_{j}) = (u_{j}) \cdot {\textrm{tanh}} [{\textrm{ln}}(1 + e^{u_{j}})] \ ,  
\end{equation}
with derivative

\begin{equation}
\label{eqn:d_mish}
g_{j}'(u_{j}) = {\textrm{sech}}^2[\textrm{softplus}(u_j)] \cdot \textrm{silu}({u_{j})} + \frac{g_{j}(u_{j})}{u_{j}} \ .  
\end{equation}

At first sight, Mish looks similar to SiLU, as can be observed in figure~\ref{fig: activation functions}. Mish is smooth, non-monotonic (bounded below and unbounded above), self-regularised and continuously differentiable to avoid singularities. The softplus function in the definition of Mish ensures that the output of the function is always greater than or equal to $-1$. In image classification and object detection tasks, Mish consistently outperforms ReLU and SiLU between 1 to 3 per cent across all standard architectures \citep{misra_mish_2019}.

\subsection{Photometric redshift performance metrics}\label{sec:metrics}

We assess the photo-$z$ results obtained from our new \textsc{annz+} using several performance metrics. This section provides the definitions for all the metrics used to quantify the performance of point estimate photo-$z$s produced after testing for each activation function. 

\begin{enumerate}
\item \textit{Root-mean-square error, $\sigma_{\textrm{RMS}}$}, 
\begin{equation}
\label{eqn:sigma_rms}
\sigma_{\textrm{RMS}}=\sqrt{\frac{1}{N} \sum_{i}^{N} \Delta z_i^2} \ .
\end{equation}
The difference between the photometric and spectroscopic redshift, scaled by $1 + z_{\textrm{spec}}$, is $\Delta z_i = (z_{\textrm{phot},i} - z_{\textrm{spec},i})/(1+z_{\textrm{spec},i})$. It is important to note that $\sigma_{\textrm{RMS}}$ is calculated without outliers removed. Therefore, it measures the overall scatter of the sample.

\item \textit{68th percentile error, $\sigma_{68}$},
\begin{equation}
\label{eqn:sigma_68}
\sigma_{\textrm{68}}=\frac{Q_{84.1\%} (\Delta z_i) - Q_{15.9\%} (\Delta z_i)}{2} \ ,
\end{equation}
is defined as the half width of the weighted distribution of $\Delta z_i$ containing $68$ per cent of the objects, where $Q$ are the quantiles of the distribution of $\Delta z_i$. This metric is centralised and assesses the width of the core in the photo-$z$ distribution, with the sensitivity to outliers greatly reduced.

\item \textit{Outlier fraction, $\eta_{out}$},\\
is the percentage of objects for which  
\begin{equation}
\label{eqn:out_frac}
\left| \Delta z_i \right| \geq 0.15\ .
\end{equation}
This metric \citep{ilbert_nout_2006} identifies the percentage of objects with large outliers. 
\end{enumerate}

\section{Data}\label{sec:data}
In order to study the performance of activation functions towards the accuracy of photo-$z$s, \textsc{annz} requires two types of data in each sample: spectroscopic (true) redshifts and multi-band photometric data. The data was obtained from two surveys: SDSS (LRG and Stripe-82 are the special subsets within the larger SDSS project) and PAUS. The LRG and Stripe-82 datasets utilised five broad bands $ugriz$ along with SDSS spectroscopic redshifts. In contrast, the PAUS dataset utilised PAUS-GALFORM catalogue, which contains six broad bands ($ugr{i^*}z{y^*}$)\footnote{We note that CFHT $y^{*}$-band filter (CFHT identification: i.MP9702) is not the infrared band, it is the successor of the original CFHT $i^*$-band filter (CFHT identification: i.MP9701) which broke in 2008. Observations in the $y^{*}$-band filter were conducted for a total of 33 fields, including 19 fields that had problematic PSF properties in the original $i^*$-band observations. They decided to distinguish the two with labels $i^*$ for i.MP9701 and $y^*$ for i.MP9702 \cite{erben_2013}.} and 40 narrow bands, with spectroscopic redshifts from the simulation. The data is divided into three samples: LRG, Stripe-82 and PAUS-GALFORM samples. All samples of galaxies are used to test the activation functions to get the general results and for specific purposes: the PAUS-GALFORM sample used to test the effectiveness of the activation functions on high dimensional input training, while the LRG and Stripe-82 samples are used for photo-$z$ comparison with other works. 

\begin{figure}
    \includegraphics[width=\linewidth]{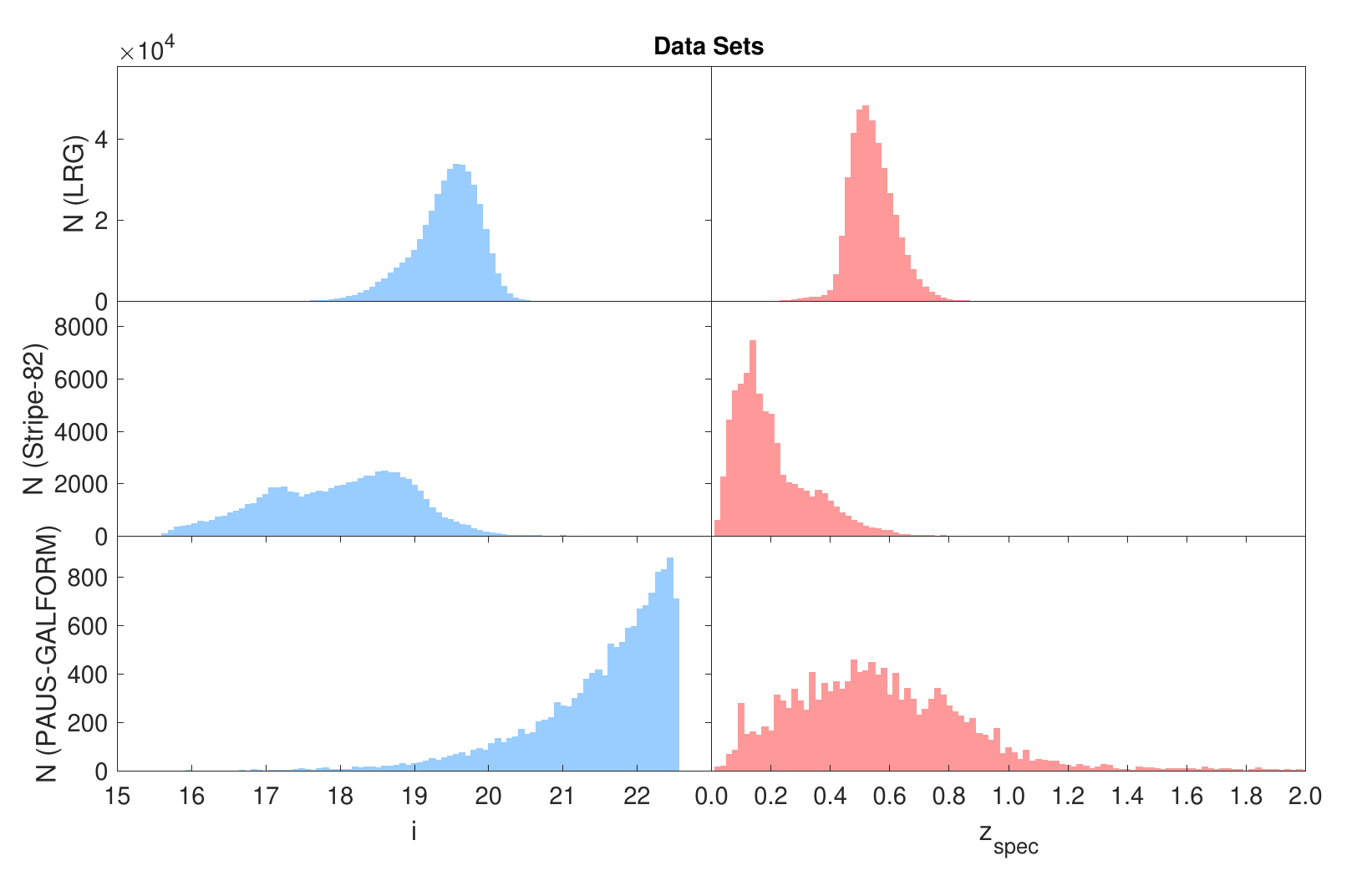}
    \caption{Distribution of $i$-band magnitude (blue histogram) and spectroscopic redshift (red histogram) for the LRG, Stripe-82 and PAUS-GALFORM samples.}
    \label{fig:data}
\end{figure}

We define the testing set as the set of data without redshift information, where the aim is to determine the photo-$z$s; the performance metrics in eqs.~(\ref{eqn:sigma_rms}),~(\ref{eqn:sigma_68}) and~(\ref{eqn:out_frac}) are evaluated using this set; the training set consists of available and reliable spec-$z$s information of galaxies; while the validation set is used as part of the training process to prevent overtraining. The total number of galaxies in the samples is divided equally into a 1:1:1 ratio of training-validating-testing sets as suggested by \cite{soo_morpho-z_2018}. Sections \ref{sec:lrg}, \ref{sec:stripe82} and \ref{sec:PAUS} describe the sources of the data for this study as well as the training samples used in this paper. The distribution of galaxies is illustrated in figure~\ref{fig:data}.

\subsection{SDSS LRG Sample}\label{sec:lrg}
The SDSS luminous red galaxy (hereafter, LRG) \cite{eisenstein_lrg_2001} sample is selected using colour and magnitude; the LRGs are characterised by their red colour, high luminosity and faint apparent magnitude due to the presence of the strong 4000 \AA\:break despite lower S/N. LRGs are typically massive elliptical galaxies at intermediate redshifts ($0.4 \le z \le 0.7$). The LRGs provided the clearest early observation of baryon acoustic oscillations \citep{eisenstein_BOA_2005}. The MegaZ-LRG catalogue \citep{collister_megaz-lrg_2007,abdalla_comparison_2011} is an example of an LRG catalogue that relied on photo-$z$. 

We applied photometric cuts as detailed in \cite{eisenstein_lrg_2001} and cross-match with photometry from the SDSS Catalog Archive Server Jobs System (CasJobs) DR12 \citep{alam_eleventh_2015}. We obtained spectroscopic data for the LRG sample through the Legacy Survey also via SDSS CasJobs. To do this, we used the flag \texttt{TARGET=GALAXY\textunderscore RED} and removed redshifts of low quality by setting \texttt{zWarning=0}. We then selected objects with redshifts between $0.1<z<1$, \texttt{zErr>0}, and magnitudes between $20.0<u<28.0$, $19.5<g<24.5$, $18.0<r<22.2$, $17.3<i<20.9$, and $16.7<z<21.0$. This generated a sample of 412,138 galaxies.

\subsection{SDSS Stripe-82 Sample}\label{sec:stripe82}
The SDSS Stripe-82 Survey \cite{annis_sloan_2014} is a compilation of imaging data (co-addition) \cite{jiang_sloan_2014}, obtained by repeatedly scanning along the equatorial stripe with the following coordinates: $-50^\circ \leq$ RA $\leq 60^\circ$ and $-1.25^\circ \leq$ DEC $\leq 1.25^\circ$. The survey reaches approximately two magnitudes fainter than the standard SDSS observations, with an approximate magnitude limit of $i\sim24.1$. We opted to work with objects in the Stripe-82 region because SDSS provide photometric data and plenty of spectroscopic redshift data. The wide-angle deep imaging data from this region has been widely used in various studies, such as photo-$z$ estimation \citep{reis_sloan_2012,soo_morpho-z_2018,alshuali_photometric_2022}, quasar classification \citep{peters_quasar_2015}, massive galaxy evolution \citep{bundy_galaxy_2015}, the creation of deeper co-adds \citep{fliri_stripe82_2015}, radio emission in radio-quiet quasars \citep{liao_radio_2022} and active galactic nuclei \citep{carvajal_agn_2023}. 

The photometric data of this area can be obtained from the SDSS CasJobs, with the setting \texttt{run=106,206}. It is important to note that any runs other than these are either data prior to co-addition or were not observed under photometric conditions. The following are the selection criteria for this sample: for the photometry, only objects from \texttt{run=106,206} (co-added photometry), the galaxies with redshift between $0<z<1$, magnitudes is between $17.0<u<31.0$, $16.0<g<26.0$, $16.0<r<24.5$, $15.6<i<24.0$ and $15.2<z<24.0$ are used. Outlier and bad data has been removed from the sample, based on the following criteria: $\delta u<5.0$, $\delta g<1.0$, $\delta r<0.3$, $\delta i<0.3$, and $\delta z<1.0$. 

The spectroscopic data in this sample use $z_{\textrm{spec}}$ within the Stripe-82 region from SDSS spectrograph \cite{york_sloan_2000} which have $9 \times 10^{5}$ galaxies with Petrosian magnitude $r\sim17.77$. To ensure only quality redshifts are used, we select all redshifts that have \texttt{zWarning=0} and require spectral \texttt{class=="GALAXY"}. We have utilised the SDSS star-galaxy separator \texttt{TYPE=3} for our sample, which helped us filter out non-extended objects. After applying these selection criteria, a sample of 72,763 galaxies was obtained.

\subsection{PAUS-GALFORM Sample}\label{sec:PAUS}
The Physics of the Accelerating Universe Survey (hereafter, PAUS) is a narrowband photometric galaxy survey which used the 4.2 m William Herschel Telescope (WHT) at the Roque de los Muchachos
Observatory (ORM) on the Canary island of La Palma, Spain \citep{benítez_baryon_2009,castander_pau_2011}. PAUS employed the PAUCam instrument using 40 narrowband filters ranging between 4500 and 8500 \AA\: with a spacing of 100 \AA, complemented with six standard broad bands \cite{padilla_paucam_2019}. The survey has successfully observed over 51 ${\textrm{deg}}^2$ of the sky since the beginning of 2016.
When using the \textsc{bcnz2}, it is possible to achieve $\sigma_{68}/(1 + z) = 0.0037$ for $i_{\text{AB}} < 22.5$ in the redshift range $0 < z < 1.2$, when selecting the best 50 per cent of the sources based on a photometric redshift quality cut \cite{eriksen_pau_2019}.

Here, we utilise the synthetic dataset from the PAUS \texttt{GALFORM} mock catalogue \cite{Manzoni2024}, which consists of 14,053 galaxies. The dataset was simulated to mimic the observational characteristics of PAUS using the \texttt{GALFORM} semi-analytical model of galaxy formation within the Planck Millennium N-body \cite{baugh_planck_2019}. By simulating the galaxy population to produce realistic properties such as positions, luminosities, colours, and redshifts, the PAUS-GALFORM mock catalogue serves as a valuable tool for studying galaxy evolution, validating models, and gaining insights into the physical processes shaping galaxies in the Universe. The lightcone mock catalogue is built by estimating galaxy brightness and positions by interpolating smoothly between output redshift from the Planck Millennium N-body simulation, using the point at which a galaxy crosses the observer's lightcone (see ref. \cite{Merson2013, Stothert2018} for further details). The mock catalogue for PAUS covers approximately 100 ${\textrm{deg}}^2$ and is magnitude limited to $i=22.5$, covering a redshift range $0<z<2$.

In our case, we did not use any photometric errors from \cite{cabayol_error_2021}, nor did we introduce any errors in the model galaxy photometry, as done by \cite{Manzoni2024},  as we do not want the errors to affect the performance of the ANN alone. However, we acknowledge that the photo-$z$ generated from the PAUS-GALFORM sample is constructed based on a limited number of initial parameters and synthetic magnitudes, this idealisation could allow \textsc{annz+} to learn the underlying parameterisation of the semi-analytic model. This is unlike observed data such as the SDSS LRG and Stripe-82 samples, which have added real-world variability. By leveraging both mock and observed datasets, we provide a balanced assessment that highlights the robustness of the activation functions across different data scenarios and ensure that the conclusions drawn from the mock data are not overly dependent on its idealised characteristics. The photometric data has been bifurcated into two distinct samples: the PAUS-GALFORM(6), hereafter referred to as the PG-06 sample, which uses six broad bands $ugr{i^*}z{y^*}$ and the PAUS-GALFORM(46), we label as the PG-46 sample, which utilises both broad bands and the 40 narrow bands. The six bands will be used to compare the performance of the activation functions along with LRG and Stripe-82 samples, while 46 bands were used to test the impact of the activation function on high dimensional inputs. 


\section{Impact of activation functions on photometric redshift performance}\label{sec:results}

In this section, our primary focus is not on obtaining the absolute global minimum value with the best performance metrics because the most common approach to achieve the absolute global minimum is through trial and error since specific hyperparameters may not work for every sample. Instead, we are more interested in ensuring the consistency and stability of \textsc{annz+} under any condition and making it suitable for universal use. We have tested various activation functions in \textsc{annz+} to observe their behaviour on different samples, various network architectures, random seed variations, and the effectiveness of each activation function. 

\subsection{General results}\label{sec:general}

\begin{figure}
    \includegraphics[width=\linewidth]{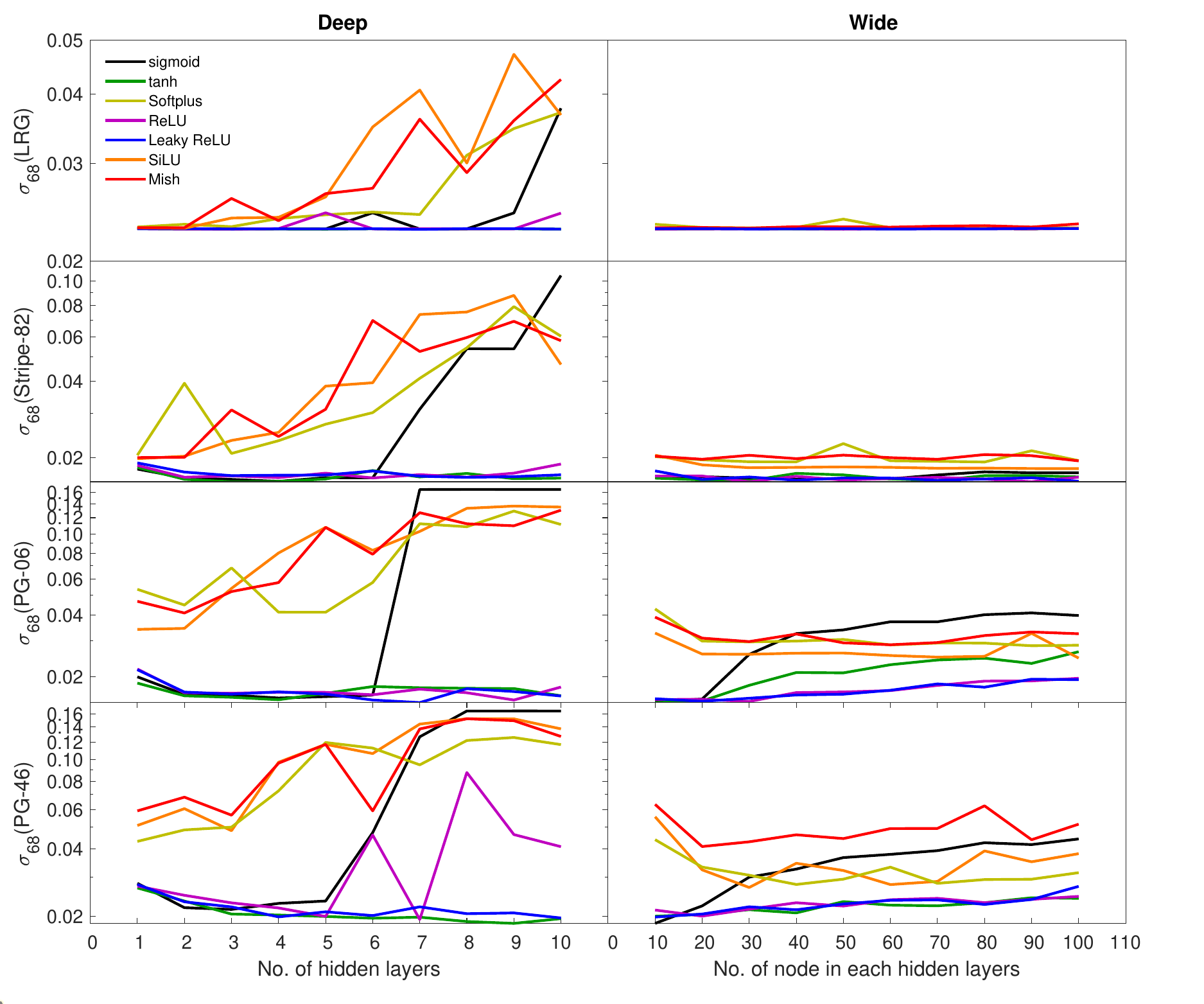}
    \caption{Relation between the number of hidden layers (deep network) as well as the number of nodes in each hidden layer (wide network) using LRG, Stripe-82, PG-06 and PG-46 samples corresponds to the $\sigma_{\textrm{68}}$ for different activation functions as indicated by different colour lines. The pattern of the line plot was used to determine the performance of each activation function. Tanh, Leaky ReLU and ReLU consistently have the lowest performance metric, which is considered the best activation function.}
    \label{fig: sig68 vs hiddenlayer}
\end{figure}

\begin{table}
\begin{tabular}{l l rrrrrr}        
\hline
\textbf{Architecture} & \textbf{Functions} & \textbf{$\sigma_{\textrm{RMS}}$} & \textbf{$\Delta\%$} & \textbf{$\sigma_{\textrm{68}}$} & \textbf{$\Delta\%$} & \textbf{$\eta_{\textrm{out}} (\%)$} & \textbf{$\Delta\%$}\\
\hline     
\multicolumn{8}{c}{\textbf{LRG}}\\
\hline
5:10:10:1 & sigmoid & 0.02811 &  & 0.02285 &  & 0.18 &  \\
5:30:30:1 & tanh & 0.02810 & 0.1 & 0.02285 & 0.0 & 0.18 & 1.6 \\
5:40:40:1 & tanh & 0.02811 & 0.0 & 0.02285 & 0.0 & 0.18 & -0.4 \\
5:20:20:1 & sigmoid & 0.02811 & 0.0 & 0.02285 & 0.0 & 0.18 & 0.8 \\
5:10:10:1 & tanh & 0.02812 & 0.0 & 0.02285 & 0.0 & 0.18 & -0.4 \\
5:40:40:1 & Leaky ReLU & 0.02812 & 0.0 & 0.02285 & 0.0 & 0.18 & 2.0 \\
5:10:10:1 & tanh & 0.02812 & 0.0 & 0.02285 & 0.0 & 0.18 & -0.4\\
\hline
\multicolumn{8}{c}{\textbf{Stripe-82}}\\
\hline
5:10:10:1 & sigmoid & 0.02591 &  & 0.01662 &  & 0.33  &  \\
5:10:10:1 & tanh & 0.02569 & 0.8 & 0.01641 & 1.3 & 0.33  & 0.0 \\
5:$10_{4}$:1 & tanh & 0.02573 & 0.7 & 0.01609 & 3.2 & 0.33  & 1.2 \\
5:20:20:1 & tanh & 0.02582 & 0.3 & 0.01615 & 2.8 & 0.37  & -12.5 \\
5:30:30:1 & tanh & 0.02594 & -0.1 & 0.01603 & 3.5 & 0.36  & -8.8 \\
5:100:100:1 & Leaky ReLU & 0.02657 & -2.6 & 0.01595 & 4.0 & 0.40  & -20.0 \\
5:30:30:1 & ReLU & 0.02659 & -2.6 & 0.01588 & 4.5 & 0.35  & -5.0 \\
\hline
\multicolumn{8}{c}{\textbf{PG-06}} \\
\hline
6:10:10:1 & sigmoid & 0.03365 &  & 0.01635 &  & 0.64  &  \\
6:$10_{10}$:1 & tanh & 0.03005 & 10.7 & 0.01612 & 1.4 & 0.57 & 11.1 \\
6:$10_{3}$:1 & tanh & 0.03020 & 10.2 & 0.01585 & 3.1 & 0.59  & 7.4 \\
6:$10_{4}$:1 & sigmoid & 0.03123 & 7.2 & 0.01572 & 3.9 & 0.71  & -11.1 \\
6:$10_{4}$:1 & tanh & 0.03154 & 6.3 & 0.01542 & 5.7 & 0.71  & -11.1 \\
6:$10_{9}$:1 & ReLU & 0.03176 & 5.6 & 0.01537 & 6.0 & 0.74  & -14.8 \\
6:$10_{7}$:1 & Leaky ReLU & 0.03292 & 2.2 & 0.01493 & 8.7 & 0.67  & -3.7 \\
\hline
\multicolumn{8}{c}{\textbf{PG-06 vs PG-46}} \\
\hline
6:10:10:1 & sigmoid & 0.03365 &  & 0.01635 &  & 0.64  &  \\
46:$10_{8}$:1 & tanh & 0.03231 & 4.0 & 0.01896 & -16.0 & 0.64 & 0.0 \\
46:$10_{10}$:1 & tanh & 0.03238 & 3.8 & 0.01954 & -19.5 & 0.71 & -11.1 \\
46:$10_{7}$:1 & tanh & 0.03249 & 3.5 & 0.01983 & -21.2 & 0.67 & -3.7 \\
46:$10_{6}$:1 & tanh & 0.03298 & 2.0 & 0.01962 & -19.9 & 0.67 & -3.7 \\
46:$10_{5}$:1 & tanh & 0.03321 & 1.3 & 0.01998 & -22.1 & 0.67 & -3.7 \\
46:$10_{4}$:1 & Leaky ReLU & 0.03326 & 1.2 & 0.01988 & -21.6 & 0.67 & -3.7 \\
\hline
\multicolumn{8}{c}{\textbf{PG-46}} \\
\hline      
46:10:10:1 & sigmoid & 0.03532 &  & 0.02186 &  & 0.74 &  \\
46:$10_{8}$:1 & tanh & 0.03231 & 8.5 & 0.01896 & 13.3 & 0.64 & 12.9 \\
46:$10_{10}$:1 & tanh & 0.03238 & 8.3 & 0.01954 & 10.6 & 0.71 & 3.2 \\
46:$10_{7}$:1 & tanh & 0.03249 & 8.0 & 0.01983 & 9.3 & 0.67 & 9.7 \\
46:$10_{6}$:1 & tanh & 0.03298 & 6.6 & 0.01962 & 10.3 & 0.67 & 9.7 \\
46:$10_{4}$:1 & Leaky ReLU & 0.03326 & 5.8 & 0.01988 & 9.0 & 0.67 & 9.7 \\
46:$10_{9}$:1 & tanh & 0.03335 & 5.6 & 0.01860 & 14.9 & 0.74 & 0.0 \\
\hline
\end{tabular}
\centering
\caption{Relative improvement (indicate by $\Delta\%$) in root-mean-square error ($\sigma_{\textrm{RMS}}$), 68th percentile error ($\sigma_{\textrm{68}}$) and outlier fraction $\eta_{\textrm{out}}$ for the LRG, Stripe-82 and PAUS-GALFORM (PG) samples, with respect to architecture 5:10:10:1 and sigmoid activation function. The table below only shows the top 6 results, where we see that the tanh, ReLU, and Leaky ReLU activation functions have been shown to produce competitive results. PG-06 utilises the $ugr{i^*}z{y^*}$ broad bands while PG-46 incorporates both broad bands and 40 narrow bands.}  
\centering
\label{tab:res}
\end{table}

Figure~\ref{fig: sig68 vs hiddenlayer} shows our results when testing deep and wide networks for the LRG, Stripe-82, and PAUS-GALFORM samples. The 68th percentile error for tanh, ReLU, and Leaky ReLU consistently demonstrated stable trends for all the samples, whether deep (more layers) or wide (more neurons or nodes per layer). When we delve into a comparison between sigmoid, Mish, softplus, and SiLU in a deep network, it becomes apparent that the error increases significantly as we deepen the network architecture, i.e. as we add more "hidden" layers. On the other hand, increasing the number of nodes per layer gives us a contrasting outcome, with the results turning out to be more stable than the deep networks. However, the processing time for wide networks is long. For instance, the computational runtime required to process the largest sample used in this work (i.e. the LRG sample) was 150 minutes for the 5:$10_{5}$:1 architecture, while it took 547 minutes for the 5:50:50:1 architecture. The LRG sample contains 412,138 objects, leading to a mean runtime per object for each architecture to be approximately 0.02 seconds and 0.08 seconds, respectively. In the context of high-dimensional input data, the PG-46 sample took a maximum runtime of 36 minutes for deep networks and 150 minutes for wide networks, utilising the mentioned architecture. The average runtime per object, across a total of 14,053 objects was 0.15 seconds (deep) and 0.64 seconds (wide). Consequently, the increased number of inputs resulted in a longer average runtime per object.

After testing with several activation functions, we found that the 5:10:10:1 architecture is enough for the $\tanh$ function to achieve the same convergence as the sigmoid function on the LRG sample. A complex architecture is unnecessary for the LRG sample. Adding extra hidden layers to a network architecture leads to greater performance gains compared to widening existing layers \cite{firth_estimating_2003}; however, for any data set, there will be a fundamental limit to this gain: this limit is determined by the inherent variation in the data set, representing the best possible accuracy that can be achieved with the given data. The required network complexity and size of the training set will depend on this limit, as well as other factors, such as the signal-to-noise ratio. 

Even though the sigmoid function can still provide a reasonable result when using simple networks, problems arise when dealing with deep networks and high-dimensional input (see figure~\ref{fig: sig68 vs hiddenlayer}). We will explain in more detail about high-dimensional input in section~\ref{sec:dimensional}. As we can see in figure~\ref{fig: sig68 vs hiddenlayer}, the sigmoid function's performance degrades from the 7th layer onwards for the Stripe-82 and PAUS-GALFORM samples. The function reaches saturation region (as shown in figure~\ref{fig: activation functions}) at the larger and smaller input, causing a smaller output range between 0 and 1. As the backpropagation process continues, the gradients progressively shrink until they vanish, leaving the initial hidden layers without gradient information. This is known as the vanishing and exploding gradient problem, and it is difficult for neural networks to learn and optimise effectively. Thus, it is evident that \textsc{annz} will need to be updated with more stable activation functions.

After analysing the overall situation, we narrowed our focus in table~\ref{tab:res} to examine the individual activation functions in more detail. The table summarises the relative improvement for the top 6 architectures with lowest values of $\sigma_{\textrm{RMS}}$, $\sigma_{\textrm{68}}$ and outlier fractions for LRG, Stripe-82 and PAUS-GALFORM samples. The $\Delta\%$ indicates the improvement with respect to the performance metric of default setting: architecture, $N_{\textrm{in}}:10:10:1$ and sigmoid activation functions. Upon conducting tests on the LRG samples, we observed that the values for all metrics using tanh, Leaky ReLU, ReLU and sigmoid with different architectures were consistent with the default setting. We did not notice any significant improvement due to the proper selection cut of the data, which is the redshift range is small, luminosity and colour of the objects are similar. Overall, not much variety in this dataset.

Regarding the Stripe-82 sample, it can be concluded that using tanh does not improve $\sigma_\textrm{RMS}$ by more than 1 per cent. However, it is worth noting that tanh has a correlation with improving 68th percentile error, showing an improvement of 1.3 to 3.2 per cent. Meanwhile, by widening the existing layers, 4.5 per cent improvements of $\sigma_{\textrm{68}}$ can be achieved when using ReLU, followed by Leaky ReLU (4.0 per cent). These points suggest that tanh, ReLU, and Leaky ReLU are suitable for the $\sigma_{\textrm{68}}$ performance metric in wider networks.

In the PG-06 sample, using tanh, sigmoid, ReLU and Leaky ReLU significantly improved the results. We observed a notable correlation of the metrics by utilising the tanh activation function with three hidden layers. This resulted in a 10.2 per cent of $\sigma_{\textrm{RMS}}$, a 3.1 per cent $\sigma_{\textrm{68}}$, and a 7.4 per cent improvement in the outlier fraction. The activation functions ReLU and Leaky ReLU have been found to significantly improve the accuracy of $\sigma_{\textrm{68}}$ in this sample, with a maximum improvement of 8.7 per cent. This finding is consistent with the trend observed with the Stripe-82 sample, highlighting the effectiveness of these activation functions in reducing the sensitivity to outliers in Stripe-82 and PAUS-GALFORM samples. Due to the lack of galaxies at higher redshift, it is necessary to use a deep network or add more hidden layers to the ANNs to improve the learning process.

On the other hand, we found that it is not advisable to use Mish, SiLU and softplus activation functions in photo-$z$ estimation due to their poor performance, despite their ability to outperform ReLU in other problems. In this particular study, we found out that no particular architecture will work for every sample indicating that the use of the most recently published activation function is not necessarily the best choice. 

\subsection{Sensitivity to random seed}\label{sec:optimisation}

\begin{figure}
    \includegraphics[width=\linewidth]{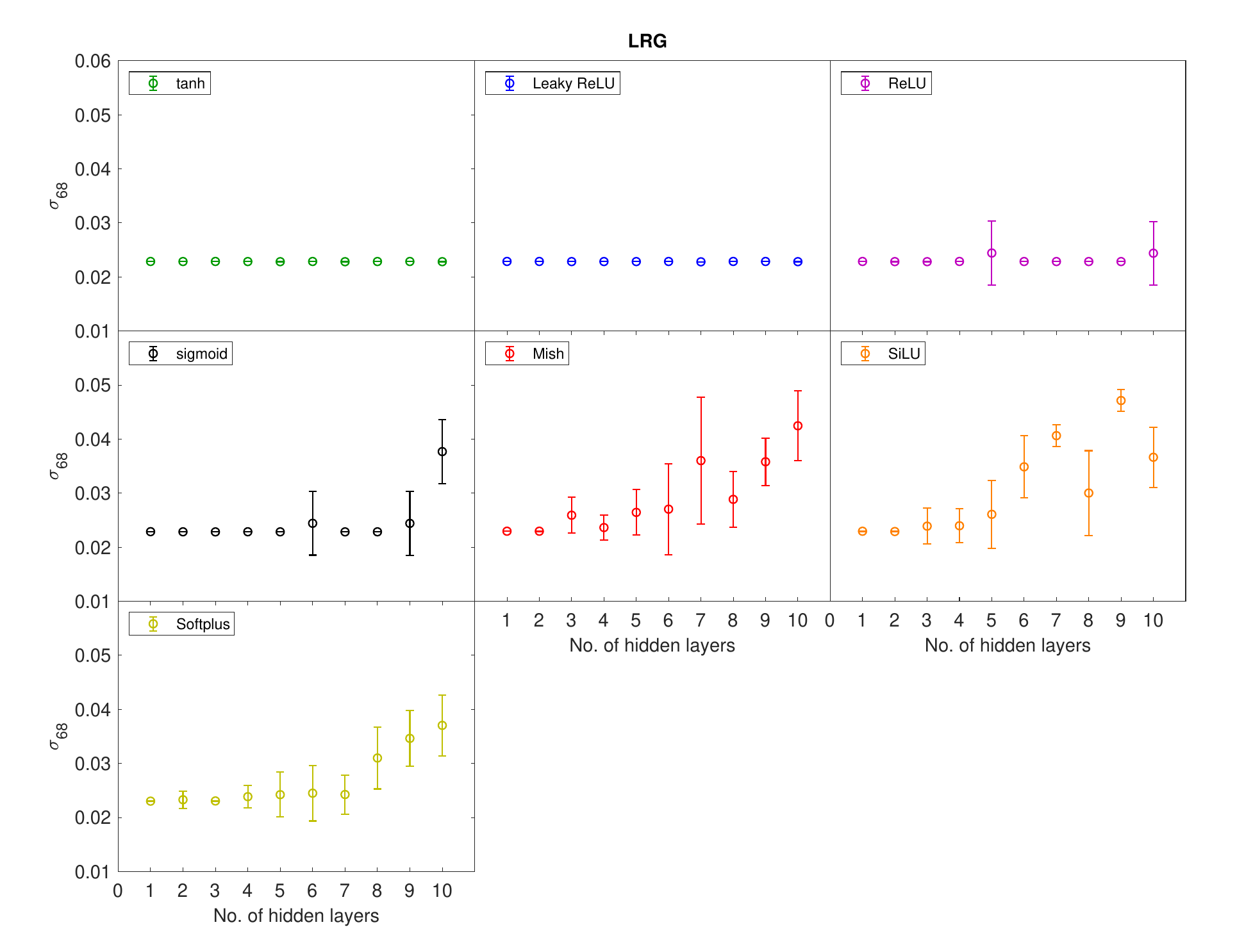}
    \caption{A comparison of seven activation functions—tanh, Leaky ReLU, ReLU, sigmoid, Mish, SiLU, and softplus—across varying numbers of hidden layers for the LRG sample. Each point represents the $\sigma_{\textrm{68}}$ value for a specific activation function with error bars. As the model complexity increases, we observe distinct trends for each function.}
    \label{fig: sig68_lrg}
\end{figure}

\begin{figure}
    \includegraphics[width=\linewidth]{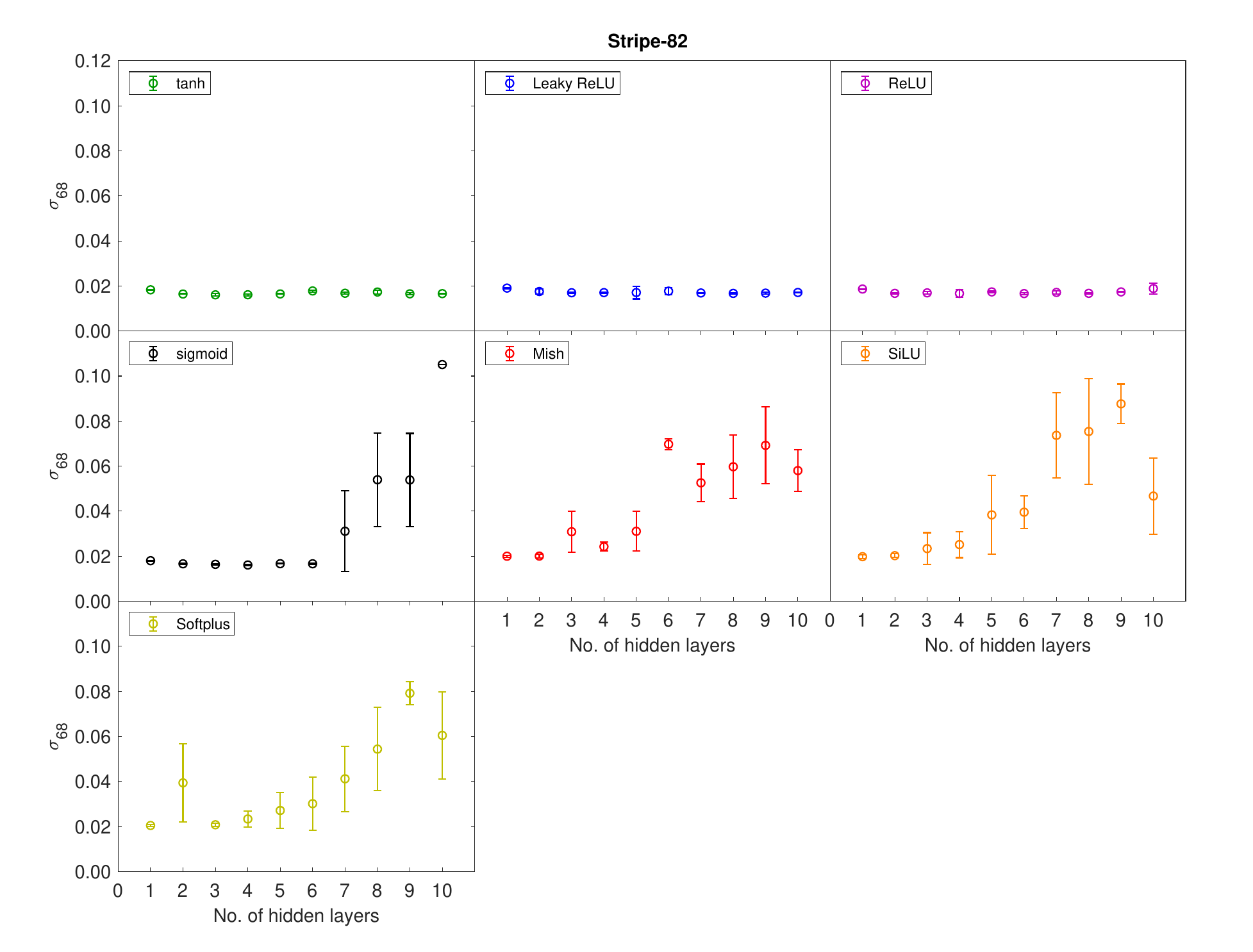}
    \caption{A comparison of seven activation functions—tanh, Leaky ReLU, ReLU, sigmoid, Mish, SiLU, and softplus—across varying numbers of hidden layers for the Stripe-82 sample. Each point represents the $\sigma_{\textrm{68}}$ value for a specific activation function with error bars. As the model complexity increases, we observe distinct trends for each function.}
    \label{fig: sig68_s82}
\end{figure}

\begin{figure}
    \includegraphics[width=\linewidth]{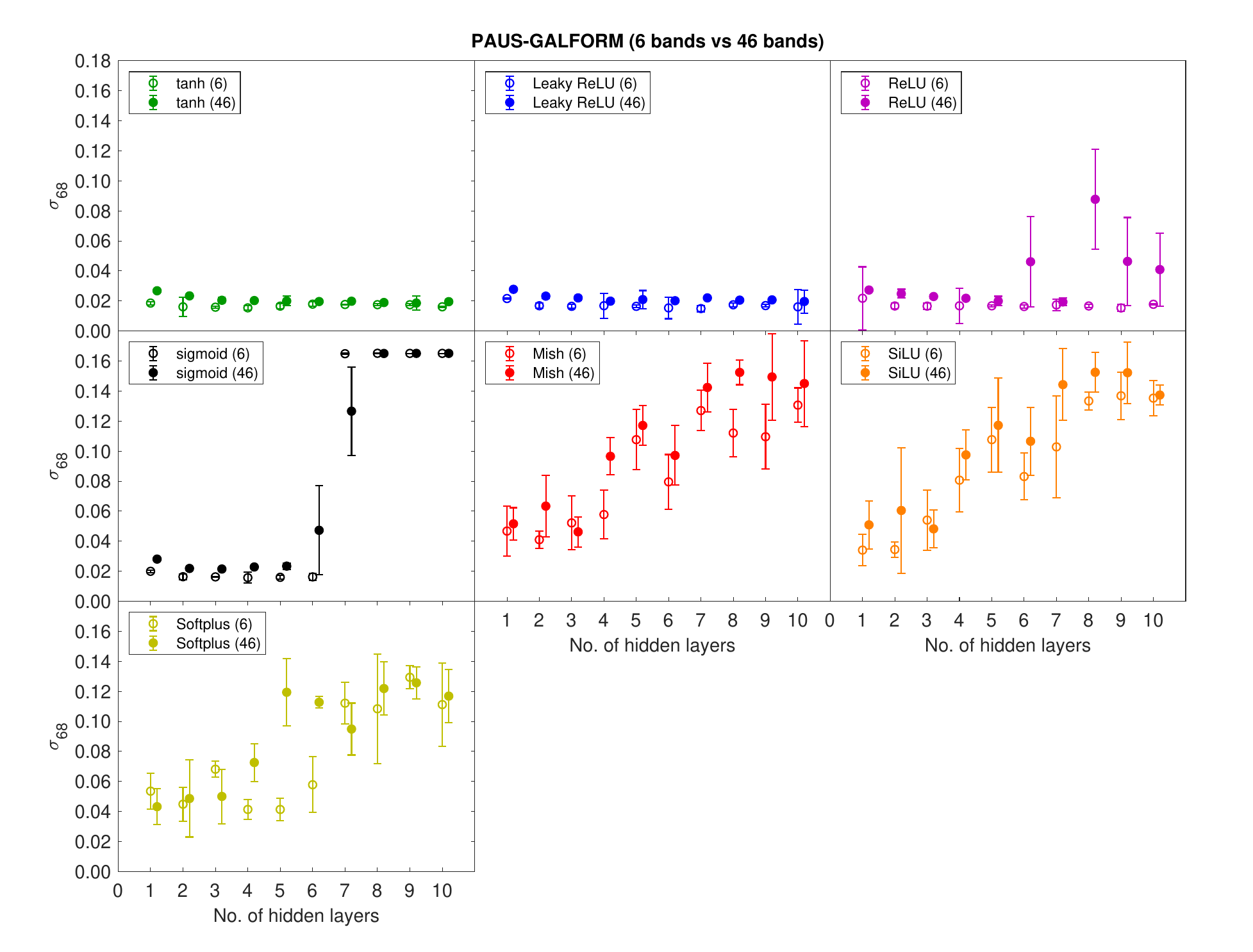}
    \caption{A comparison of seven activation functions—tanh, Leaky ReLU, ReLU, sigmoid, Mish, SiLU, and softplus—across varying numbers of hidden layers for the PAUS with six $ugr{i^*}z{y^*}$ broad bands sample. Each point represents the $\sigma_{\textrm{68}}$ value for a specific activation function with error bars. As the model complexity increases, we observe distinct trends for each function.}
    \label{fig: sig68_paus6VSpaus46}
\end{figure}

Figures~\ref{fig: sig68_lrg}, ~\ref{fig: sig68_s82}, and~\ref{fig: sig68_paus6VSpaus46} show that each activation function has a different level of sensitivity towards the choice of the random seed that determines the weight initialisation through these three samples represented by error bars. The error bars denote the standard error, which is derived from the mean of $\sigma_{\textrm{68}}$ across four committees (i.e. four choices of random seed) for each architecture. Therefore, the error bars indicate the extent of fluctuation in random seed for specific activation functions.    

Across all samples, the optimisation of tanh and Leaky ReLU is not influenced by the random seed. Meanwhile, Mish, SiLU and softplus are very sensitive; thus, without a proper selection of random seed, the error may go up by 89 per cent because of the outliers when we cycle through the SiLU, softplus and Mish activation functions. Therefore in certain runs, we tweaked and changed some of the values of random seeds in order to maintain consistent results across each set of 4 committees as sometimes they can't converge because of the exploding gradient. The default random seeds have to be changed for the PG-46 sample with 46 bands as well, due to its high number of inputs. Particularly for wide and very deep networks, the running for softplus and Mish stops at a certain number of nodes, and some runs could never complete due to the random seed that caused them to be unable to converge. Therefore, we advise users to always test out different random seeds every time the activation functions are changed to avoid problems arising when running the ANN.

Theoretically, the hyperbolic tangent function does not solve the vanishing and exploding gradient problems, but in \textsc{annz+}, it works rather well. Despite several studies claiming that SiLU, Mish, and softplus may solve the exploding gradient problem, they end up not converging in some of our runs. The ReLU and sigmoid activation functions were found to be least sensitive to random seeds, and only a few runs showed significant error bars. For sigmoid, ReLU, Mish, softplus and SiLU, a proper selection of random seed is needed in ANN to ensure fast convergence. If the initial weights are too small, then the signal flowing through the network gradually diminishes, eventually leading to a loss of information and potentially causing the network fail to learn effectively; if the weights are initialised with values too large, the variance of input data into each layer increases rapidly and soon overflows, leading to unstable training \cite{Cui_NIR_2018}.
 
After analysing the three different samples with varying galaxy distributions and redshift ranges, we can conclude that the fluctuation of error bars for the mean of the four committees, as shown in figures~\ref{fig: sig68_lrg}, ~\ref{fig: sig68_s82}, and~\ref{fig: sig68_paus6VSpaus46} remained consistent and have the same patterns for all samples. We related this pattern to the landscape of the loss function. The loss landscape\footnote{\url{https://losslandscape.com/faq/}} is the geometric landscape formed by the loss function or cost function. Each activation function has a distinct characteristic that impacts the shape and structure of the loss landscape. A comparison between the loss landscape of ReLU and Mish from the previous study reveals that ReLU functions have sharp minima, which are characterised by steep and narrow regions around the minimum \cite{misra_mish_2019}. ReLU introduces sparsity by setting negative values to zero, which can influence the density of the gradients in the loss landscape. On the other hand, Mish has a wider minimum, which results in a smoother and flatter loss landscape surface which contributes to better generalisation. Flat minima may lead to slower convergence during optimisation, as the gradients around flat minima are less informative and may require more iterations for optimisation.
 
In the context of softplus and SiLU functions, the loss landscape visualisation may contain multiple local minima, saddle points, plateaus, and steep valleys, which makes it more sensitive to the random seed and difficult to optimise. Tanh and Leaky ReLU are similar to ReLU but have sharper, steeper gradients in the shape of the loss landscape, which makes the convergence continuously differentiable across their domains. This indicates a more direct path to the global minimum of the loss function.

\subsection{Time efficiency}

\begin{table}
    \centering
    \begin{tabular}{lcrr}
    \hline
         \textbf{Functions} & \textbf{Runtime (minutes)} &\textbf{$\sigma_{\textrm{RMS}}$} & \textbf{$\sigma_{\textrm{68}}$} \\
    \hline
         sigmoid& 43 & 0 .02811 & 0.02285\\
         tanh& 56 & 0.02812 & 0.02285\\
         SiLU & 32 & 0.02822 & 0.02290\\
         Softplus & 37 & 0.02864 & 0.02329\\
         ReLU & 27 & 0.02814 & 0.02283\\
         Mish & 60 & 0.02841 & 0.02296\\
         Leaky ReLU & 27 & 0.02814 & 0.02287\\
    \hline
    \end{tabular}
    \caption{Comparison of runtime for deriving photo-$z$ estimates from a LRG sample containing 412,138 galaxies, alongside the performance metrics ($\sigma_{\textrm{RMS}}$ and $\sigma_{\textrm{68}}$) for various activation functions using a 5:10:10:1 network architecture.}
    \label{tab:runtime}
\end{table}

In terms of effectiveness, ReLU and Leaky ReLU are the most friendly functions with respect to time, volume of data and computational efficiency. According to table~\ref{tab:runtime}, these simple functions can converge faster and run all of the training data (sample sizes up to 400 000) in 27 minutes with competitive $\sigma_{\textrm{RMS}}$ and $\sigma_{\textrm{68}}$ values. Meanwhile, networks with sigmoid and tanh activation functions converge more slowly with respective runtimes of 43 and 56 minutes for simple architecture. A longer runtime is expected when the complexity of the network increases, especially in deeper and wider networks. Mish took the longest runtime to converge with a total time of 60 minutes due to its complex functional forms, and unfortunately, also produced the most problematic results, followed by SiLU and Softplus. In our case, we need a faster point estimator that can converge faster during optimisation. It is confirmed that sharp minima are beneficial in terms of optimisation speed and efficiency \cite{dinh_sharpminima_2017}. This point gives credit to the use of ReLU and Leaky ReLU activation functions for managing extensive data sets associated with large-scale surveys like \textit{Euclid} and LSST, which aim to calculate photo-$z$ for billions of objects.

\subsection{Comparison between ReLU variants}

Here, the tanh function is shown to be robust and is suggested to be the best activation to be used for regression problems like estimating photo-$z$s. While the ReLU activation function is widely used in neural networks due to its effectiveness in handling vanishing gradient problems and accelerating convergence, we find that it is not always necessary to use the ReLU activation function, particularly in regression analysis. The choice of activation function should be carefully considered depending on the specific requirements of the model and the objectives of the analysis.

To a certain extent, ReLU experiences the dying ReLU problem, in which ReLU completely blocks any input less than zero. In some of our training runs, the bias becomes very negative, hence causing the neurons to fail to learn during backpropagation. The inputs are not positive enough to overcome the bias and lead to the creation of many dead neurons. To solve this problem, we suggest the use of Leaky ReLU, ELU or GeLU. In our study, we can see that Leaky ReLU slightly outperforms ReLU in $\sigma_{\textrm{RMS}}$ and $\sigma_{\textrm{68}}$ , as shown in table~\ref{tab: PAUS46_AF}.

\subsection{Performance of activation functions for high-dimensional inputs}\label{sec:dimensional} 

\begin{table}
\begin{tabular}{l rrrrrr}     
\hline
\textbf{Function} & \textbf{$\sigma_{\textrm{RMS}}$} & \textbf{$\Delta\%$} & \textbf{$\sigma_{\textrm{68}}$} & \textbf{$\Delta\%$} & \textbf{$\eta_{\textrm{out}} (\%)$} & \textbf{$\Delta\%$} \\
\hline
sigmoid & 0.1594 &  & 0.1650 &  & 33.43
&  \\
\hline
tanh & 0.0323 & 79.7 & 0.0190 & 88.5 & 0.64 & 98.1 \\
Leaky ReLU & 0.0335 & 78.9 & 0.0206 & 87.5 & 0.67 & 98.0 \\
ReLU & 0.0867 & 45.6 & 0.0877 & 46.9 & 7.28 & 78.2 \\
Softplus & 0.1239 & 22.3 & 0.1219 & 26.1 & 20.96 & 37.3 \\
SiLU & 0.1541 & 3.3 & 0.1525 & 7.6 & 32.43 & 3.0 \\
Mish & 0.1572 & 1.3 & 0.1524 & 7.6 & 31.83 & 4.8 \\
\hline
\end{tabular}
\centering
\caption{A comparison of the performance of different activation functions—tanh, SiLU, softplus, ReLU, Mish, and Leaky ReLU with respect to sigmoid for photo-$z$s on the PAUS-GALFORM data with eight hidden layers, trained with 46 bands.}  
\centering
\label{tab: PAUS46_AF}
\end{table}

\begin{figure}
    \includegraphics[width=\linewidth]{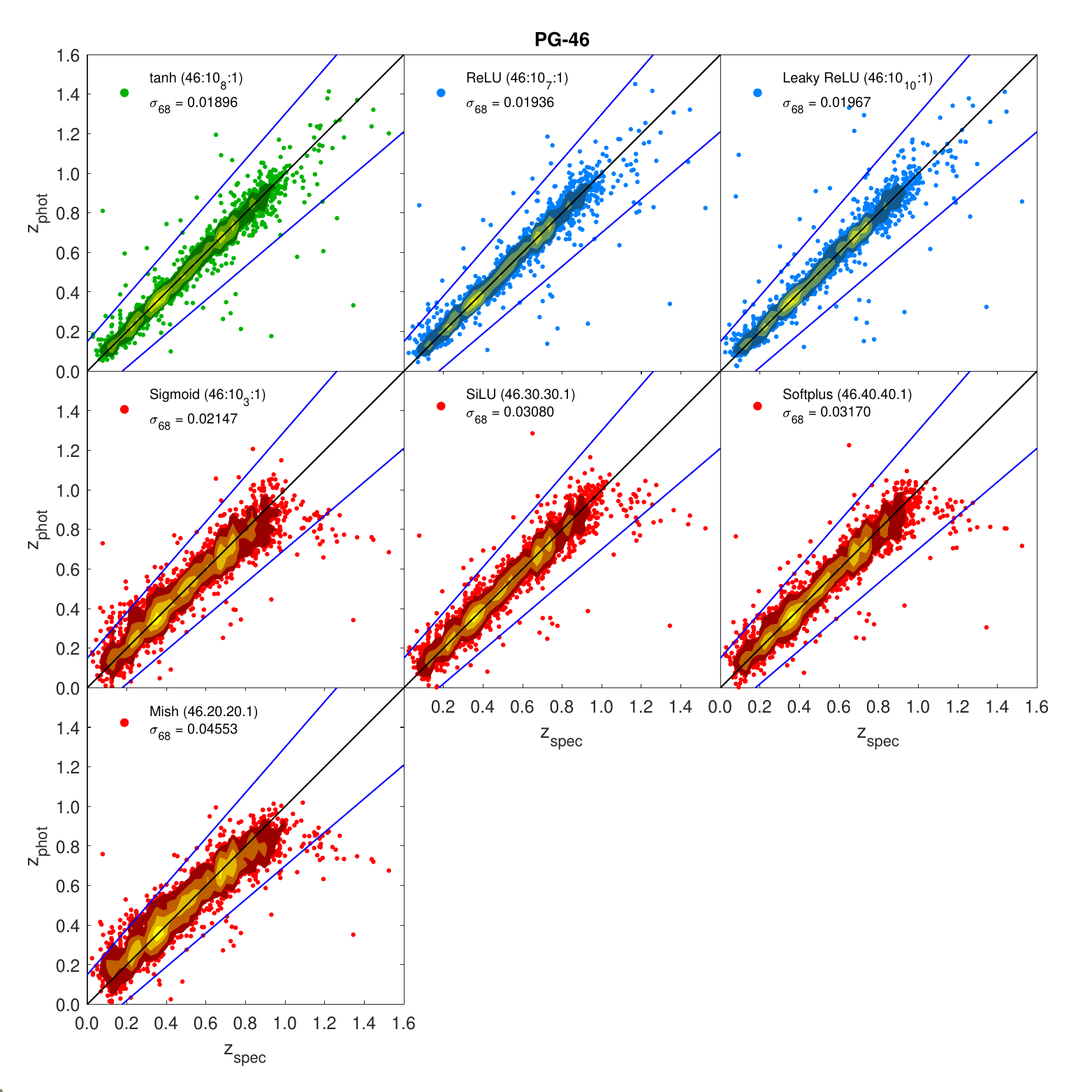}
    \caption{Plots of photo-$z$ versus spec-$z$, comparing the photo-$z$ performance when using tanh, ReLU, Leaky ReLU, sigmoid, SiLU, softplus and Mish activation functions in \textsc{annz} for the PG-46 sample (trained and tested using 6 $ugr{i^*}z{y^*}$ broad bands and 40 narrow bands). Comparing activation functions based on the lowest $\sigma_{\textrm{68}}$ for each activation function across various architectures to evaluate the performance. A deep network, denoted as $N_{\textrm{in}}:10_{d}:1$, where $N_{\textrm{in}}$ is the number of input and $d$ is the number of hidden layers. As seen from the plots, tanh, ReLU and Leaky ReLU functions showed promising results when producing photo-$z$s with a large number of inputs. The blue lines are the limit for outliers}
    \label{fig: 46inputs}
\end{figure}

The PAUS-GALFORM mock galaxy sample is an excellent sample to test our implementation of different activation functions for the photo-$z$s in a high dimensional space since it consists of 40 narrow bands and six broadband filters. We vary the choice of activation functions to gain valuable insights into how different activation functions respond to the complexity and high dimensionality of the PAUS data. From table~\ref{tab:res}, we see that the tanh function in a deep network when testing with 46 bands brings about up to 4.0 per cent improvement in $\sigma_{\textrm{RMS}}$ to photo-$z$ with respect to the PG-06 sample with the default setting of \textsc{annz}. This indicates that the tanh function is helpful in enabling ANNs to estimate photo-$z$ from 46 bands. This improvement can potentially solve the issue raised in ref. \cite{collister_annz_2004}, where additional input could lead to a decline in the performance of ANNs.

Upon comparison of performance between the PG-46 sample with respect to the default setting, it was observed that utilising deeper architecture and the tanh activation function resulted in a notable improvement of up to 8.5 per cent in $\sigma_{\textrm{RMS}}$, 14.9 per cent in $\sigma_{\textrm{68}}$ and 12.9 per cent improvement of the outlier fraction. By adding more hidden layers, ReLU and Leaky ReLU also achieved better performance. This indicates that changing the activation function and utilising deeper architecture can help to reduce the impact of the high dimensionality and achieve significant improvements in performance than changing the architecture alone. This could solve the problem as mentioned in ref. \citep{soo_pau_2021}.

It was found that deep networks potentially performed well for the PAUS data, therefore, we fixed the architecture to 46:10:10:10:10:10:10:10:10:1 to assess the impact of deep network architecture on each activation function. Table~\ref{tab: PAUS46_AF} tabulates the results of each activation function's performance on this sample with respect to the sigmoid activation function. The results show that using tanh activation functions in high-dimensional inputs and deep networks can lead to a 79.7 per cent improvement in $\sigma_{\textrm{RMS}}$, 88.5 per cent in $\sigma_{\textrm{68}}$ and 98.1 per cent in the outlier fraction compared to sigmoid, followed by Leaky ReLU. The difference in $\sigma_{\textrm{RMS}}$ (34.1 per cent) and $\sigma_{\textrm{68}}$ (41.6 per cent) between tanh and ReLU proves that ReLU is not always the best even though ReLU is the most popular activation function in the literature. Despite the improvements found with softplus, SiLU, and Mish, the performance metrics exhibit a significant range (0.1219 to 0.1572), and the presence of an outlier fraction which deviates from the norm. 

It is evident that the sigmoid function failed and did not facilitate effective learning in the ANNs, as it did yield the highest performance metric. This is due to the derivative of the sigmoid activation function becomes saturated (close to zero) for large positive and large negative inputs (see figure~\ref{fig: activation functions}). From eq.~(\ref{eqn:cost}), the derivative of the error function concerning the weights and biases is proportional to the product of the derivatives of the activation functions of each layer. Taking an eight-layer network with sigmoid activation as an example, the derivative of the error function with respect to the weight $w_{ij}$ connecting the $i$-th neuron in the seven-layer to the $j$-th neuron in the eight-layer is
\begin{equation}
    \frac{\partial{E}}{\partial{w_{ij}^{(7)}}} 	\propto g'[u_{j}^{(8)}]~g'[u_{i}^{(7)}] \ ,
\end{equation}
where $u_{j}^{(8)}$ and $u_{i}^{(7)}$ are the weighted inputs of the neurons on the eighth and seventh layers, respectively. If the values of $u_{j}^{(8)}$ and $u_{i}^{(7)}$ are very large or very small, then $g'[u_{j}^{(8)}]~g'[u_{i}^{(7)}]$ will be very close to zero (see sigmoid activation function derivatives in figure~\ref{fig: activation functions}), and so will $\frac{\partial{E}}{\partial{w_{ij}^{(7)}}}$. This means that the weight $w_{ij}^{(7)}$ will not be updated significantly during the backpropagation algorithm, and the network will not learn effectively. The ANNs converge before the limit and are unable to improve beyond that. The same problem applies to biases as well. This illustrates the vanishing gradient phenomenon, which makes the learning process very slow and can prevent the neural network from reaching the optimal weights and biases.

The scatter plot in figure~\ref{fig: 46inputs} displays the correlation between photo-$z$ and spec-$z$ with the diagonal line representing perfect correlation.  Comparing activation functions based on the lowest $\sigma_{\textrm{68}}$ for each activation function across various architectures to evaluate the performance. Tanh aligns closely along the diagonal, approaching accurate predictions. ReLU and Leaky ReLU also shows good alignment, but the galaxies scatter more at higher redshifts. Sigmoid, SiLU, softplus and Mish exhibit mediocre performance. This statistical analysis demonstrates that using sigmoid, softplus, SiLU and Mish with high-dimensional inputs is not recommended. These photo-$z$ results will be compared with \textsc{delight}, \textsc{bcnz2} and \textsc{cigale} in an upcoming paper (Soo et al. in prep).

\section{Application: performance comparison with other works}  

\subsection{Comparison of ANNZ+ with ANNZ2}

\begin{table}
\tiny
\begin{tabular}{l rrrrrrrrrrrr}             
\hline
 & \multicolumn{4}{c}{\textbf{Default}} & \multicolumn{4}{c}{\textbf{Deep}} & \multicolumn{4}{c}{\textbf{Wide}} \\
\hline
\textbf{Photo-z estimator} & $\sigma_{\textrm{RMS}}$ & $\Delta\%$ & $\sigma_{\textrm{68}}$ & $\Delta\%$ & $\sigma_{\textrm{RMS}}$ & $\Delta\%$ & $\sigma_{\textrm{68}}$ & $\Delta\%$ & $\sigma_{\textrm{RMS}}$ & $\Delta\%$ & $\sigma_{\textrm{68}}$ & $\Delta\%$  \\
\hline
\multicolumn{13}{c}{\textbf{LRG}}\\
\hline
\textsc{annz2} (sigmoid) & 0.0289 &  & 0.0232 &  & 0.0290 &  & 0.0232 &  & 0.0288 &  & 0.0232 &  \\
\textsc{annz} (sigmoid) & 0.0281 & 2.7 & 0.0228 & 1.4 & 0.0410 & -41.4 & 0.0377 & -62.8 & 0.0281 & 2.4 & 0.0229 & 1.4 \\
\textsc{annz2} (tanh) & 0.0288 &  & 0.0231 &  & 0.0290 &  & 0.0232 &  & 0.0288 &  & 0.0231 &  \\
\textsc{annz+} (tanh) & 0.0281 & 2.5 & 0.0228 & 1.1 & 0.0281 & 3.1 & 0.0228 & 1.8 & 0.0281 & 2.4 & 0.0229 & 1.2 \\
\hline
\multicolumn{13}{c}{\textbf{Stripe-82}}\\
\hline
\textsc{annz2} (sigmoid) & 0.0288 &  & 0.0189 &  & 0.0284 &  & 0.0191 &  & 0.0289 &  & 0.0187 &  \\
\textsc{annz} (sigmoid) & 0.0259 & 10.1 & 0.0166 & 11.8 & 0.0989 & -248.7 & 0.1051 & -451.2 & 0.0264 & 8.5 & 0.0164 & 12.5 \\
\textsc{annz2} (tanh) & 0.0279 &  & 0.0187 &  & 0.0290 &  & 0.0189 &  & 0.0281 &  & 0.0183 &  \\
\textsc{annz+} (tanh) & 0.0257 & 7.8 & 0.0164 & 12.0 & 0.0260 & 10.2 & 0.0166 & 12.1 & 0.0266 & 5.4 
& 0.0169 & 7.6 \\
\hline
\multicolumn{13}{c}{\textbf{PAUS-GALFORM (6)}}\\
\hline
\textsc{annz2} (sigmoid) & 0.0419 &  & 0.0278 &  & 0.0438 &  & 0.0325 &  & 0.0413 &  & 0.0275 &  \\
\textsc{annz} (sigmoid) & 0.0337 & 19.7 & 0.0164 & 41.2 & 0.1595 & -264.0 & 0.1651 & -408.5 & 0.0480 & -16.2 & 0.0357 & -29.8 \\
\textsc{annz2} (tanh) & 0.0491 &  & 0.0261 &  & 0.0390 &  & 0.0257 &  & 0.0631 &  & 0.0250 &  \\
\textsc{annz+} (tanh) & 0.0345 & 29.8 & 0.0161 & 38.3 & 0.0300 & 22.9 & 0.0161 & 37.2 & 0.0375 & 40.6 & 0.0223 & 10.9 \\
\hline
\multicolumn{13}{c}{\textbf{PAUS-GALFORM (46)}} \\
\hline
\textsc{annz2} (sigmoid) & 0.0443 &  & 0.0338 &  & 0.0475 &  & 0.0368 &  & 0.0493 &  & 0.0360 &  \\
\textsc{annz} (sigmoid) & 0.0353 & 20.3 & 0.0219 & 35.4 & 0.1595 & -235.9 & 0.1651 & -348.7 & 0.0540 & -9.7 & 0.0410 & -13.7 \\
\textsc{annz2} (tanh) & 0.0443 &  & 0.0333 &  & 0.0459 &  & 0.0346 &  & 0.0585 &  & 0.0333 &  \\
\textsc{annz+} (tanh) & 0.0363 & 18.1 & 0.0234 & 29.8 & 0.0324 & 29.5 & 0.0195 & 43.6 & 0.0393 & 32.8 & 0.0269 & 19.1 \\
\hline
\end{tabular}
\centering
\caption{A comparison of three photo-z estimators—\textsc{annz} (sigmoid), \textsc{annz2} (sigmoid and tanh), and \textsc{annz+} (tanh)—across different neural network architectures: default, deep, and wide. The table includes improvement percentage $\sigma_{\textrm{RMS}}$ and $\sigma_{\textrm{68}}$ for each estimator on LRG, Stripe-82 and PAUS-GALFORM samples. The performance of \textsc{ANNz+} shows a significant improvement compared to other estimators.}  
\centering
\label{tab:annz_comparison}
\end{table}

\begin{figure}
    \includegraphics[width=\linewidth]{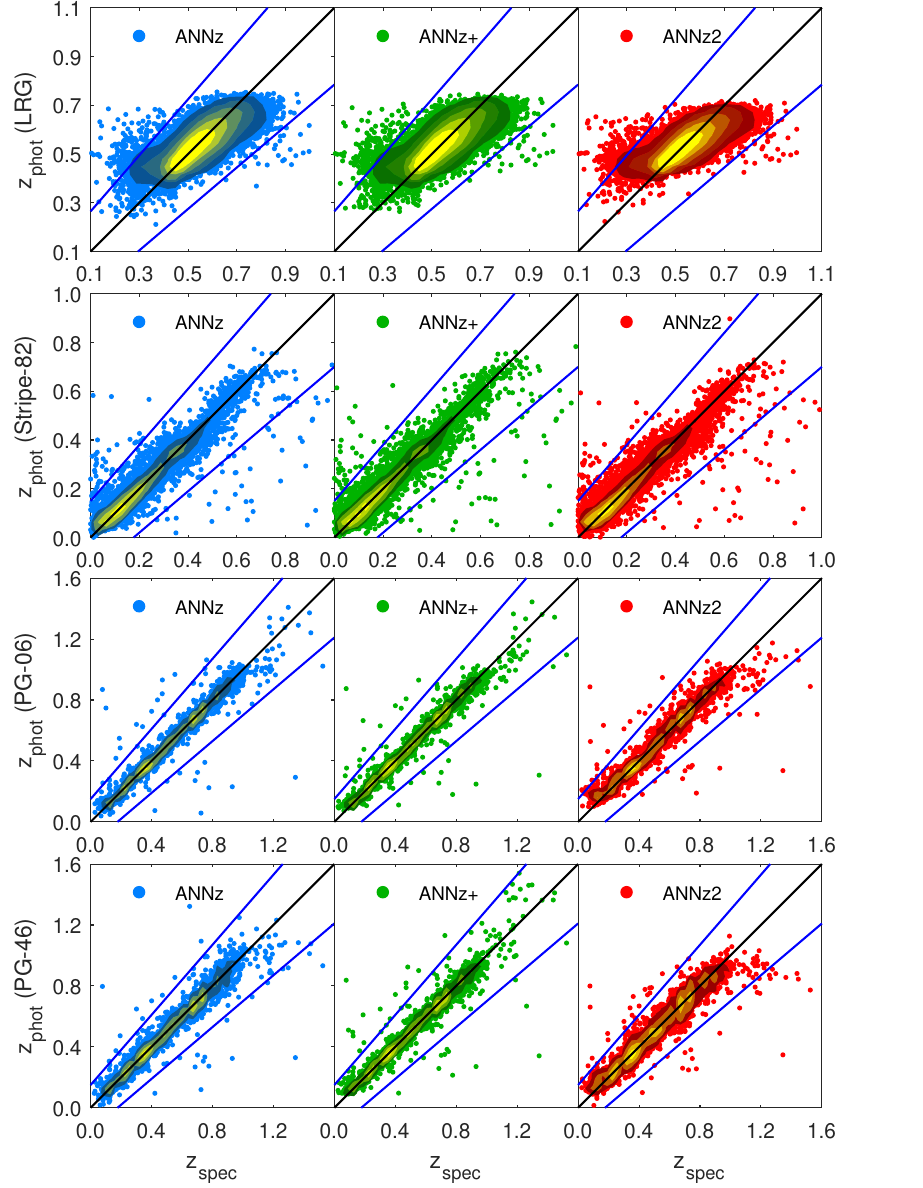}
    \caption{Plots of photo-z versus spec-z, comparing the results of the $\sigma_{\textrm{68}}$ of \textsc{annz} (blue), \textsc{annz+} (green) and \textsc{annz2} (red) for all samples.}
    \label{fig:zphot_zspec}
\end{figure}

We are interested in comparing the performance of \textsc{annz+} and \textsc{annz2} when the ANN hyperparameters are fixed, such as sample sizes, network architecture, and activation functions. \textsc{annz2} \cite{sadeh_annz2_2016} is the successor to \textsc{annz}, which differs in programming language and functionality despite their similar names. The \textsc{annz2} algorithm is a powerful machine learning based photo-$z$ estimator that uses a mixture of different machine learning methods, including artificial neural networks, boosted decision trees, and $k$-nearest neighbours, to produce probability density functions (PDFs) that provide highly accurate photo-$z$ point estimates. This algorithm uses the Toolkit for Multivariate Data Analysis (\textsc{tmva}) \cite{hoecker_tmva_2007} with \textsc{root} \cite{brun_root_1997}, which allows it to run multiple different machine learning algorithms for training and output photo-$z$s based on a weighted average of their performance. \textsc{annz2} calculate uncertainties using the $k$-Nearest Neighbors ($k$NNs) method and has been widely used in recent works \citep{alshuali_photometric_2022,schmidt_lsst_2020,soo_morpho-z_2018} due to its high customisability and ability to produce accurate results.

Here, \textsc{annz2} version 2.3.1 was run using artificial neural networks individually optimised for best performance. The configurations used were \texttt{NCycles=9999}, \texttt{EstimatorType=MSE}, \texttt{TrainingMethod=BFGS}, \texttt{LearningRate=0.02} and \texttt{UseRegulator=False}. We evaluated the performance of \textsc{annz2} using the same set of samples as presented in section~\ref{sec:data}. \textsc{annz2} comes with both sigmoid and tanh activation functions, thus we used \textsc{annz} (utilised sigmoid function) and \textsc{annz+} (utilised tanh function) to be compared with \textsc{annz2}'s results. To this end, we considered three distinct network architectures, which are $N_{\textrm{in}}:10:10:1$ (default), $N_{\textrm{in}}:10:10:10:10:10:10:10:10:10:10:1$ (deep) and $N_{\textrm{in}}:50:50:1$ (wide). For LRG and Stripe-82 samples, there are $N_{\textrm{in}} = 5$ inputs, while for PAUS samples, $N_{\textrm{in}} = 6$ and $46$ inputs were used. 

The mean value of the PDF ($z_{\textrm{pdf}}$) was chosen to represent the photo-$z$ point estimate for \textsc{annz2}, which is calculated via a weighted average across all runs. This is because ref. \cite{soo_morpho-z_2018} found that $z_{\textrm{pdf}}$ produces better overall accuracy at the high redshift end as compared to the peak (mode) value of the PDF ($z_{\textrm{peak}}$). However, if the PDF has a low redshift contribution, we acknowledge that $z_{\textrm{peak}}$ may represent the most suitable value that could yield different results from \textsc{annz2}. Moreover, regardless of the point estimate method chosen, whether the mean or mode—\textsc{annz+} consistently outperforms \textsc{annz2}. It achieves a 3.6 per cent improvement in the LRG sample, a 9.5 per cent in the Stripe-82 sample, and a 25.9 per cent in the PG-46 sample. These comparisons remain valid for \textsc{annz2} when it employs results generated from $z_{\textrm{peak}}$, utilised tanh activation function and adheres to the default architecture. The purpose of this investigation was to obtain insight into the impact of network architecture on the overall performance of \textsc{annz2}. 

The improvement percentages of photo-$z$ estimation results of these runs for \textsc{annz} and \textsc{annz+} as compared to \textsc{annz2} are summarised in table~\ref{tab:annz_comparison} and visualised in the figure~\ref{fig:zphot_zspec}. The overall result shows \textsc{annz+} performs much better than \textsc{annz2}. As we can see, the performance metrics of \textsc{annz2} on $\sigma_{\textrm{RMS}}$ and $\sigma_{\textrm{68}}$ are higher than \textsc{annz+}, resulting in poorer performance of \textsc{annz2} compared to \textsc{annz+} with difference up to 2.5 per cent for the LRG sample, 12.1 per cent for Stripe-82, 40.6 per cent for the PG-06 sample and 43.6 per cent difference for the PG-46 sample despite changes in the activation function. Surprisingly, even though \textsc{annz2} uses the sigmoid function, it still works well in deep networks compared to the original \textsc{annz}, which gives worse results. Furthermore, the use of tanh has been shown to enhance the accuracy of photo-$z$s in \textsc{annz2}. Figure~\ref{fig:zphot_zspec}, shows \textsc{annz+} gets many of the high spec-$z$ objects correctly close to the diagonal line of the scatter plot for Stripe-82, PG-06 and PG-46. This shows that \textsc{annz+}, despite showing not much difference in results for the LRG sample, has shown promising results to produce better photo-$z$s for high redshift samples. As we move to the \textsc{annz2}, it is evident that \textsc{annz2} does not perform well when dealing with high dimensional inputs where the scatter plot density of galaxies did not focus on the regression line.

The discrepancies between both algorithms are due to the regularisation method, backpropagation uncertainty generation method and the optimisation process in between \textsc{annz} and \textsc{annz2}. Ref. \cite{soo_pau_2021} showed that \textsc{annz2} performed poorly in high-dimensional input when a large number of inputs is used, which aligns with our findings. The utilisation of $k$NNs by \textsc{annz2} to generate errors might be the underlying reason, it looks at the nearby objects in the significantly high-dimensional $ugr{i^*}z{y^*} + 40$ narrowband space and uses them to determine the associated uncertainties. We can see that the $k$NN method probably works the best when the number of inputs is small, as the curse of dimensionality may have diluted the separating power of $k$NNs to determine better uncertainties. Despite differences in architecture, it is worth noting that \textsc{annz+} demonstrate a higher level of stability than their predecessors. We also recognise that \textsc{annz2} was designed to generate accurate PDFs, thus it may not be optimised for producing derived point estimates. These observations provide strong justification for updating \textsc{annz} to ensure that it remains relevant and competitive alongside \textsc{annz2}.

\subsection{Crossmatch with MegaZ-LRG catalogue and Stripe-82 Survey photometric redshifts}

\begin{table}
\begin{tabular}{l l lll}  
\hline
\textbf{Method} & \textbf{Reference} & \textbf{$\sigma_{\textrm{RMS}}$} & \textbf{$\sigma_{\textrm{68}}$} & \textbf{$\eta_{\textrm{out}} (\%)$}\\
\hline
\multicolumn{5}{c}{\textbf{MegaZ-LRG Catalogue}}\\
\hline
\textsc{annz} & \cite{collister_annz_2004} & 0.0266 & 0.0231 & 0.04 \\
\hline
\textsc{annz+} (tanh) & This work & 0.0263 & 0.0226 & 0.05 \\
\textsc{annz2 x bpz} & \cite{alshuali_photometric_2022} & 0.0265 & 0.0222 & 0.07 \\
\textsc{annz2} (tanh) & \cite{sadeh_annz2_2016} & 0.0267 & 0.0229 & 0.04 \\
SDSS template & SDSS pipeline & 0.0426 & 0.0286 & 0.99 \\
\textsc{zebra} & \cite{feldmann_zurich_2006} & 0.0472 & 0.0325 & 1.35 \\
\textsc{bpz} & \cite{benitez_bayesian_2000} & 0.0495 & 0.0387 & 1.55 \\
\textsc{hyperz cww} & \cite{bolzonella_photometric_2000} & 0.0513 & 0.0292 & 1.79 \\
\textsc{hyperz bc} & \cite{bolzonella_photometric_2000} & 0.0523 & 0.0409 & 2.45 \\
\textsc{lephare} & \cite{arnouts_measuring_1999} & 0.0371 & 0.0253 & 0.20 \\
\hline
\multicolumn{5}{c}{\textbf{Stripe-82}}\\
\hline
\textsc{annz} & \cite{collister_annz_2004} & 0.0212 & 0.0155 & 0.16 \\
\hline
\textsc{annz+} (tanh) & This work & 0.0210 & 0.0149 & 0.19 \\
Morpho-$z$ & \cite{soo_morpho-z_2018} & 0.0429 & 0.0388 & 0.39 \\
\textsc{annz2} (tanh) & \cite{sadeh_annz2_2016} & 0.0236 & 0.0174 & 0.20 \\
Reis (ANN) & \cite{reis_sloan_2012} & 0.0244 & 0.0168 & 0.26 \\
\hline
\end{tabular}
\centering
\caption{Comparison of the photometric redshift performance of other works were used to compare with our results (\textsc{annz+}) for the testing set of the MegaZ-LRG catalogue and Stripe-82 survey. The performance metrics with lower $\sigma_{\textrm{RMS}}$ and $\sigma_{\textrm{68}}$ indicate better accuracy.}
\centering
\label{tab:methods}
\end{table}

We crossmatch our \textsc{annz+} (tanh) results from the LRG sample with the MegaZ-LRG catalogue \citep{abdalla_comparison_2011} and a few other works, summarising the results in table~\ref{tab:methods}. We can see that the results between \textsc{annz}, \textsc{annz2} and \textsc{annz+} are consistent up to a difference of 0.001 in $\sigma_{\textrm{RMS}}$. We note that this result is expected since the LRG sample has a very stringent cut in terms of colour, redshift range and luminosity. We have shown that the results of \textsc{annz+} remain the lowest in terms of $\sigma_{\textrm{RMS}}$ and most competitive in its results, overtaking the results of one of the most recent methods which is a combination of \textsc{annz2} and \textsc{bpz} \citep{alshuali_photometric_2022}. In term of $\sigma_{\textrm{68}}$, \textsc{annz2} x \textsc{bpz} improved the accuracy by 0.001 compared with  \textsc{annz+}. 

Finally, we evaluate the photo-$z$ performance for the Stripe-82 Survey as well. The photo-$z$ results of \textsc{annz+} outperform morpho-$z$, \textsc{annz2} using tanh and those by ref. \cite{reis_sloan_2012} for all metrics. We note however that morpho-$z$ have been weighted in accordance to the densities of the target sample and made use of spectroscopic redshift data from non-SDSS sources as well, thus its result is notably different (refer to ref. \cite{soo_morpho-z_2018} for more details)

\section{Conclusions and future work}

We aimed to improve the accuracy of \textsc{annz}, a 20-year-old photometric redshift algorithm, by selecting an activation function to optimise the artificial neural network and give the most stable and consistent results. In the context of photometric redshift estimation and regression, it can be concluded that the tanh function is the most stable and competitive activation function all the time. It consistently produces reliable results, unaffected by random seeds or different types of architectures. For the LRG sample, the tanh function yields consistent $\sigma_{\textrm{RMS}}$ and $\sigma_{\textrm{68}}$ values with the previous literature, approximately 0.0281 and 0.0228 followed by the Leaky ReLU and ReLU activation function. In the Stripe-82 sample, the tanh function improves the accuracy of $\sigma_{\textrm{68}}$ by 3.5 per cent when using a wider network, along with $< 1$ per cent improvement of the $\sigma_{\textrm{RMS}}$ using the default architecture. In the PAUS-GALFORM samples, the tanh function showed the lowest performance metrics, resulting in a 10.7 per cent improvement in $\sigma_{\textrm{RMS}}$ and a 5.7 per cent enhancement in $\sigma_{\textrm{68}}$ for the PG-06 sample, as well as an 8.5 per cent improvement in $\sigma_{\textrm{RMS}}$ and a 14.9 per cent enhancement in $\sigma_{\textrm{68}}$ for the PG-46 sample.

We also found out that the tanh function works well in photo-$z$ estimation when using high dimensional inputs by alleviating the vanishing and exploding gradient problems faced by the sigmoid function when going too deep network architectures. We got up to a 4.0 per cent improvement of $\sigma_{\textrm{RMS}}$ for PAUS-GALFORM 46 bands. This indicates that the tanh function is helpful in enabling ANNs to estimate photo-$z$ from 46 bands. When we fixed the architecture and compared the performance across different activation functions, tanh achieved a 79.7 per cent, followed by Leaky ReLU with a 78.9 per cent difference compared to the sigmoid function.   

At this point, ReLU and Leaky ReLU are suitable for wide networks and to get the absolute best for $\sigma_{\textrm{68}}$ as they can get approximately 8.7 per cent improvement on PAUS-GALFORM 6 bands and 4.0 - 4.5 per cent for Stripe-82 samples. The ReLU and Leaky ReLU activation functions facilitate an efficacious learning process, characterised by their capability to expedite the completion of the training process. On the other hand, it is not advisable to use Mish, SiLU and softplus activation functions in photo-$z$ estimation due to their poor performance in our results, despite their ability to outperform ReLU in other machine learning problems. Sigmoid, SiLU, softplus, and Mish are all very sensitive to random seeds; therefore, caution is needed when selecting random seeds. 

We introduced \textsc{annz+} as an improved version of \textsc{annz} by replacing the sigmoid activation function with the tanh activation function as default. To assess the relevancy and competitiveness of \textsc{annz+}, we compared it with \textsc{annz2} using the tanh activation function and found a maximum difference in the performance of 43.6 per cent. A similar improvement can be seen in \textsc{annz} when compared with \textsc{annz2} using a sigmoid function, except when applied to deep and wide network architecture in \textsc{annz}. We crossmatch the MegaZ-LRG catalogue and Stripe-82 Survey with the other works, \textsc{annz+} consistently outperforms the other methods. 

We recommend the photo-$z$ community to use tanh when dealing with regression analysis because most of the newest activation functions generally only do well in classification problems. To ensure optimal performance across all conditions, it is imperative to identify the most suitable activation function. This involves a thorough analysis of various factors that impact the function's efficacy, such as the type of data being processed and the specific task at hand. By carefully selecting the appropriate activation function, one can achieve optimal results and enhance the overall efficiency of the system. For normal regression analysis, we find that it is not necessary for the ANN to go too deep as it is not worth the additional training time, while if we want to get the absolute best, we can attempt going deep using the robust tanh function.

Further investigation into whether \textsc{annz+} is consistently better than \textsc{annz2} is needed. As it stands, the point photo-$z$ estimate with \textsc{annz+} is indeed better. However, upcoming surveys such as \textit{Euclid} and LSST will probe higher redshifts and the complexity of photometric redshift estimation will increase due to redshift degeneracies, such as the Lyman/Balmer break degeneracy, which becomes more prominent for samples spanning $2<z<3$. In such cases, algorithms that are capable of producing PDFs rather than point estimates are likely to perform better and thus are more valued by the community. While \textsc{annz+} currently focuses on point estimates, it might be possible to modify \textsc{annz+} to produce PDFs and see if it can compete with \textsc{annz2}. To implement PDF generation within the \textsc{annz+} framework, several modifications are necessary. First, we need to write additional C++ scripts to locate the output layer configuration, allowing for the production of either Gaussian parameters or binned probabilities. Next, we should define the loss function to align with these modifications. Finally, we will need to create additional scripts to plot the PDF. This could be a good project to explore in the future.


\acknowledgments

IMP would like to thank Joseph Ooi Boon Han and Yoon Tiem Leong, for the fruitful discussions. JYHS acknowledges financial support via the Fundamental Research Grant Scheme (FRGS) by the Malaysian Ministry of Higher Education with code FRGS/1/2023/STG07/ USM/02/14. JYHS and IMP acknowledge financial support from the Short Term Grant by Universiti Sains Malaysia with code 304/PFIZIK/6315395. IMP also acknowledges financial assistance under the Gra-Assist 2020 scheme provided by the Institute of Postgraduate Studies (IPS), Universiti Sains Malaysia.

CMB acknowledges support from the Science Technology Facilities Council through ST/X001075/1. GM is supported by the Collaborative Research Fund under Grant No. C6017-20G which is issued by the Research Grants Council of Hong Kong S.A.R. EG acknowledges grants from Spain Plan Nacional (PGC2018-102021-B-100) and Maria de Maeztu (CEX2020-001058-M). FJC acknowledges grants from Spain Plan Nacional (PID2022-141079NB-C31) and Maria de Maeztu (CEX2020-001058-M). JC acknowledges support from the grant PID2021-123012NA-C44 funded by MCIN/AEI/ 10.13039/501100011033 and by “ERDF A way of making Europe”. 

CP acknowledges support from the Spanish Plan Nacional project PID2021-123012NB-C41. JGB acknowledges grants from Spain Plan Nacional (PID2021-123012NB-C43) and Centro de Excelencia Severo Ochoa (CEX2020-001007-S). RM acknowledges support from the Spanish Plan Nacional project PID2021-123012NB-C41. CP acknowledges support from the Spanish Plan Nacional project PID2019-111317GB-C32 and PID2022-141079NB-C32. PR acknowledges the support by the Tsinghua Shui Mu Scholarship, the funding of the National Key R\&D Program of China (grant no. 2023YFA1605600), the National Science Foundation of China (grant no. 12073014 and 12350410365), the science research grants from the China Manned Space Project with No. CMS-CSST2021-A05, and the Tsinghua University Initiative Scientific Research Program (No. 20223080023). ES acknowledges grants from Spain Plan Nacional (PID2021-123012NB-C42). ISN acknowledges grants from Spain Plan Nacional (PID2021-123012NB-C42). ME acknowledges funding by MCIN with funding from European Union NextGenerationEU (PRTR-C17.I1) and by Generalitat de Catalunya.

Funding for the SDSS-I/II/III has been provided by the Alfred P. Sloan Foundation, the Participating Institutions, the National Science Foundation, the U.S. Department of Energy, the National Aeronautics and Space Administration, the Japanese Monbukagakusho, the Max Planck Society, and the Higher Education Funding Council for England. The SDSS-I/II web site is http://www.sdss.org/, while the SDSS-III web site is http://www.sdss3.org/. 

SDSS-I/II/III is managed by the Astrophysical Research Consortium for the Participating Institutions. The Participating Institutions include the American Museum of Natural History, Astrophysical Institute Potsdam, University of Basel, University of Cambridge, Case Western Reserve University, University of Chicago, Drexel University, Fermilab, the Institute for Advanced Study, the Japan Participation Group, Johns Hopkins University, the Joint Institute for Nuclear Astrophysics, the Kavli Institute for Particle Astrophysics and Cosmology, the Korean Scientist Group, the Chinese Academy of Sciences (LAMOST), Los Alamos National Laboratory, the Max-Planck-Institute for Astronomy (MPIA), the Max Planck-Institute for Astrophysics (MPA), New Mexico State University, Ohio State University, University of Pittsburgh, University of Portsmouth, Princeton University, the United States Naval Observatory, the University of Washington, University of Arizona, the Brazilian Participation Group, Brookhaven National Laboratory, Carnegie Mellon University, University of Florida, the French Participation Group, the German Participation Group, Harvard University, the Instituto de Astrofisica de Canarias, the Michigan State/Notre Dame/JINA Participation Group, Lawrence Berkeley National Laboratory, Max Planck Institute for Extraterrestrial Physics, New York University, Pennsylvania State University, the Spanish Participation Group, University of Tokyo, University of Utah, Vanderbilt University, University of Virginia, and Yale University.

PAUS is partially supported by the Ministry of Economy and Competitiveness (MINECO, grants CSD2007-00060, AYA2015-71825, ESP2017-89838, PGC2018-094773, PGC2018-102021, SEV-2016-0588, SEV-2016-0597 and MDM-2015-0509). Funding for PAUS has also been provided by Durham University (ERC StG DEGAS259586), ETH Zurich, and Leiden University (ERC StG ADULT279396). The PAU data centre is hosted by the Port d’Informaci\'{o} Cien\'{i}fica (PIC), maintained through a collaboration of CIEMAT and IFAE, with additional support from Universitat Aut\'{o}noma de Barcelona and the European Research Development Fund (ERDF).



 \bibliographystyle{JHEP}
 \bibliography{biblio.bib}






\end{document}